\documentclass[12pt,preprint]{aastex}

\shorttitle{Spectral Line Survey toward NGC 2264 CMM3}
\shortauthors{Watanabe et al.}

\begin{document}


\title{Spectral Line Survey toward Young Massive Protostar NGC~2264~CMM3 
    in the 4~mm, 3~mm, and 0.8~mm Bands}


\author{Yoshimasa~Watanabe\altaffilmark{1}, Nami~Sakai\altaffilmark{1}, Ana~L\'opez-Sepulcre\altaffilmark{1}, Ryuta~Furuya\altaffilmark{1}, Takeshi Sakai\altaffilmark{2}, Tomoya Hirota\altaffilmark{3}, Sheng-Yuan~Liu\altaffilmark{4}, Yu-Nung~Su\altaffilmark{4}}



\and
\author{Satoshi Yamamoto\altaffilmark{1}}
\email{nabe@taurus.phys.s.u-tokyo.ac.jp}


\altaffiltext{1}{Department of Physics, The University of Tokyo, 7-3-1 Hongo, Bunkyo-ku, Tokyo, 113-0033, Japan}
\altaffiltext{2}{Graduate School of Informatics and Engineering, The University of Electro-Communications, Chofu, Tokyo 182-8585, Japan}
\altaffiltext{3}{National Astronomical Observatory of Japan, Osawa, Mitaka, Tokyo 181-8588, Japan}
\altaffiltext{4}{Academia Sinica, Institute of Astronomy and Astrophysics, PO Box 23-141, Taipei 106, Taiwan}


\begin{abstract}
Spectral line survey observations are conducted toward the high-mass protostar candidate NGC~2264~CMM3 in the 4~mm, 3~mm, and 0.8~mm bands with the Nobeyama 45~m telescope and the Atacama Submillimeter Telescope Experiment (ASTE) 10~m telescope.  In total, 265 emission lines are detected in the 4~mm and 3~mm bands, and 74 emission lines in the 0.8~mm band.  As a result, 36 molecular species and 30 isotopologues are identified.  In addition to the fundamental molecular species, many emission lines of carbon-chain molecules such as HC$_5$N, C$_4$H, CCS, and C$_3$S are detected in the 4~mm and 3~mm bands.  Deuterated molecular species are also detected with relatively strong intensities.  On the other hand, emission lines of complex organic molecules such as HCOOCH$_3$, and CH$_3$OCH$_3$ are found to be weak.  For the molecules for which multiple transitions are detected, rotation temperatures are derived to be 7--33~K except for CH$_3$OH.  Emission lines with high upper-state energies ($E_{\rm u} > 150$~K) are detected for CH$_3$OH, indicating existence of a hot core.  In comparison with the chemical composition of the Orion~KL, carbon-chain molecules and deuterated molecules are found to be abundant in NGC~2264~CMM3, while sulfur-bearing species and complex organic molecules are deficient.  These characteristics indicate chemical youth of NGC~2264~CMM3 in spite of its location at the center of the cluster forming core, NGC~2264~C. 
\end{abstract}





\section{Introduction}
Spectral line surveys are of fundamental importance in astrochemistry and astrophysics.  They reveal chemical characteristics of a target source without any preconception, and the result is subject to detailed chemical and physical modeling to strengthen basic concepts of astrochemistry, as well as to explore physical conditions of the target source.  In the course of the spectral line survey, many new interstellar molecules have been discovered, and new molecular tracers which highlight particular physical situations have been recognized.  Since the early days of radioastronomy, spectral line surveys have been conducted toward various representative sources including star-forming regions \citep[e.g.][]{Blake1986, vanDishoeck1995, Schilke1997, Bergin2010, Tercero2010, Watanabe2012}, cold dark clouds \citep{Kaifu2004}, shocked regions \citep[e.g.][]{Codella2010, Sugimura2011, Yamaguchi2012},  photodissociation regions \citep[e.g.][]{Ginard2012, Cuadrado2015}, Galactic center clouds \citep[e.g.][]{Cummins1986}, and external galaxies \citep[e.g.][]{Martin2006,Watanabe2014}.  Thanks to recent advances in receiver and backend technologies, spectral line surveys become feasible even toward fainter sources, and are being carried out toward various kinds of sources not only with single-dish telescopes but also with interferometers \citep[e.g.][]{Beuther2009,Jorgensen2011,Martin2011}. 

For instance, \citet{Caux2011} conducted the spectral line survey of the low-mass protostar, IRAS 16293-2422, which is known as a `hot corino' source \citep{Cazaux2003} with the IRAM 30~m telescope,  whereas the spectral line survey of the low-mass protostar L1527, known as a warm-carbon-chain-chemistry source \citep{Sakai2008}, was carried out with the Nobeyama 45~m telescope (Sakai~et~al. in prep.).  Although these two sources are low-mass protostars in the Class 0 stage, their spectral patterns are significantly different from each other.  This result clearly established chemical diversity of low-mass protostellar cores, whose origin and future are of great interest in relation to the origin of the Solar System \citep{Sakai2008, Sakai2009,Sakai2013}.  \citet{Watanabe2012} and \cite{Lindberg2012} reported further chemical complexity due to external UV radiation in a low-mass protostar R~CrA~IRS7B.  Furthermore, the spectral line surveys of the shocked region L1157~B1 which is caused by the interaction between the molecular outflow and the ambient gas were performed by several telescopes including the Nobeyama 45~m telescope \citep{Sugimura2011, Yamaguchi2012}, the IRAM 30~m telescope \citep[e.g.][]{Codella2012,Podio2014,Mendoza2014}, and Hershel HIFI \citep[e.g.][]{Codella2010}.  It is revealed that the shocked region harbors rich organic chemistry originating from sublimation and/or sputtering of grain mantles.  This reminds us of the importance of grain-surface productions of organic molecules.

Chemical evolution of high-mass protostellar sources may be much different from that of low-mass ones, and its detailed understanding is not only important for astrochemistry but also useful for finding diagnostic tools of high-mass star formation.  With these motivations, we conducted spectral line survey observations toward the high-mass star-forming region NGC~2264~CMM3 in the 4, 3, and 0.8~mm bands with the Nobeyama 45~m and the Atacama Submillimeter Telescope Experiment (ASTE) 10~m telescopes to characterise its chemical nature.

NGC~2264~CMM3 is the second nearest high-mass star-forming region after Orion KL \citep[$d=738$~pc:][]{Kamezaki2014}.  Many submillimeter-wave continuum sources (CMM1-13) are identified around IRS1, the brightest infrared source in the NGC~2264~C region \citep{Peretto2006,Peretto2007}.  Among them,  CMM3 is the most massive one \citep[40~M$_{\odot}$:][]{Peretto2006}, and it is believed that CMM3 will evolve into a massive star of 8~M$_{\odot}$, according to the theoretical model by \citet{Maury2009}.  However, the protostar of CMM3 is deeply embedded in the protostellar core, and invisible even in the 24 $\mu$m band of \textit{Spitzer}.  Although a molecular outflow extended toward north and south directions was detected around CMM3 in the CS ($J=5-4$) emission by \citet{Schreyer1997}, its association with CMM3 was not evident.  \citet{Saruwatari2011} detected a compact bipolar outflow definitively associated with CMM3 in the CO($2-1$), and CH$_3$OH($5_0-4_0$~A$^+$) lines with SMA, whose dynamical age is as short as 140--2000~yr.  Hence, it is very likely that CMM3 is in the earliest evolutionary stage of a high-mass star formation.  Moreover \citet{Sakai2007A} detected the millimeter-wave lines of one of complex organic molecules, HCOOCH$_3$, toward this source with the Nobeyama 45~m telescope and Nobeyama Millimeter Array, and found that its distribution is offset from the continuum emission peak of CMM3 by $5''$--$10''$.  This result implies rich chemistry in CMM3 associated with activities of a young high-mass protostar.  


\section{Observation}
\subsection{Observation with Nobeyama 45~m} 
NGC~2264~CMM3 was observed in the 4~mm and 3~mm bands with the Nobeyama 45~m telescope at the Nobeyama Radio Observatory (NRO)\footnote{Nobeyama Radio Observatory is a branch of the National Astronomical Observatory of Japan, National Institutes of Natural Sciences} in May and December, 2014.  The observed position is : ($\alpha_{\rm J2000}$, $\delta_{\rm J2000}$) = ($6^{\rm h}\,41^{\rm m}\,12^{\rm s}.3$, $+09^{\circ}\,29'\,11''.9$).  Eight frequency settings were observed to cover the frequency range from 67.5 to 116.0~GHz.  The side-band separating (2SB) mixer receivers T70H/V and TZ1H/V \citep{Nakajima2013} were used as frontends with the typical system noise temperature of 140 -- 380~K.  The image rejection ratios were measured just before each observation session by applying artificial signals, and were assured to be better than 10~dB.  The beam size ranged from $22^{''}$ to $15^{''}$.  The backends were 16 SMA45 autocorrelators whose band width and frequency resolution each are 1600~MHz and 0.5~MHz, respectively.  The frequency resolution corresponds to a velocity resolution of 1.7~km~s$^{-1}$ at 90~GHz.  This resolution is enough for resolving spectral line profiles observed in NGC~2264~CMM3 ($\sim 3$~km~s$^{-1}$).  The position-switching method was employed with the off-position at ($\Delta\alpha$, $\Delta\delta$) = (-25$'$, -25$'$).  The telescope pointing was checked every hour by observing the SiO maser source SY-Mon.  Pointing accuracy was confirmed to be better than $5''$.  Intensity scale was calibrated to the antenna temperature ($T^{*}_{\rm a}$) scale by using the chopper-wheel method, and its accuracy is estimated to be 20~\%.  The antenna temperature was converted to the main-beam brightness temperature ($T_{\rm mb}$) by $T_{\rm mb}=T_{\rm A}^{*}/\eta_{\rm mb}$, where $\eta_{\rm mb}$ is 0.38, 0.31, 0.26, and 0.24 at 75, 86, 110, and 115~GHz, respectively.  The observation parameters are summarized in Table~\ref{tab0}.

The observed data were reduced by using the software package,  NEWSTAR, developed by NRO.  Spectral baselines were subtracted by fitting a 5th - 7th order polynomial to the line-free part of spectra in a frequency range of $\sim$1.5~GHz.  Distorted sub-scan spectra due to bad atmospheric conditions and instabilities of the receiver system, whose baseline could not be subtracted by the polynomial fitting, were excluded in the integration procedure.

\subsection{Observation with ASTE} 
Observations with the ASTE 10~m telescope \citep{eza2004} were carried out in May and November 2011.  The observed position is the same as that in the observations with the Nobeyama 45~m telescope.  Eighteen frequency settings were observed to cover the frequency range from 330 to 366~GHz.  In the 345~GHz band, the beam size is $\sim 22''$, and the main beam efficiency ($\eta_{\rm mb}$) is $\sim 0.6$.  The side-band separating (2SB) mixer receiver CATS345 \citep{Inoue2008} was used as a frontend, whose typical system temperature ranged from 150 to 400~K, depending on the atmospheric conditions.  The backend was a bank of XF-type digital spectro-correlators MAC \citep{Sorai2000}, whose bandwidth is 512 MHz each, all having 1024 spectral channels.  The frequency resolution is 0.5~MHz, which corresponds to $\sim$0.5~km~s$^{-1}$ at 345~GHz.  The position-switching method was employed with the off-position at ($\Delta\alpha$, $\Delta\delta$) = (-25$'$, -25$'$).  The telescope pointing was checked every hour by observing the bright point-like $^{12}$CO($J=3-2$) source GX Mon.  The pointing accuracy was ensured to be better than $3''$.  The intensity calibration was carried out by the chopper-wheel method, and its accuracy is estimated to be $\sim20$~\%.  The antenna temperature ($T_{\rm A}^{*}$) was converted to the main-beam brightness temperature ($T_{\rm mb}$) by $T_{\rm mb}=T_{\rm A}^{*}/\eta_{\rm mb}$. The observation parameters are summarized in Table~\ref{tab0}. 

The observed data were reduced by using NEWSTAR.  Spectral baselines were subtracted by fitting a 5th order polynomial to the line-free part in a frequency range of 512~MHz.  Distorted sub-scan spectra were excluded in the integration procedure, as described in Section 2.1.

In addition to the line survey, we conducted a 5-point observation toward NGC~2264~CMM3 in the 342.7~GHz in order to investigate a distribution of CH$_3$OH around the protostar.  The grid spacings are $20''$, which is close to the beam size of $22''$.  The result of this 5-point observation will be shown in the Section 4.1.  

\section{Result and Analysis} 
\subsection{Detected Molecules} 
\begin{figure}
\includegraphics[angle=90,scale=0.35]{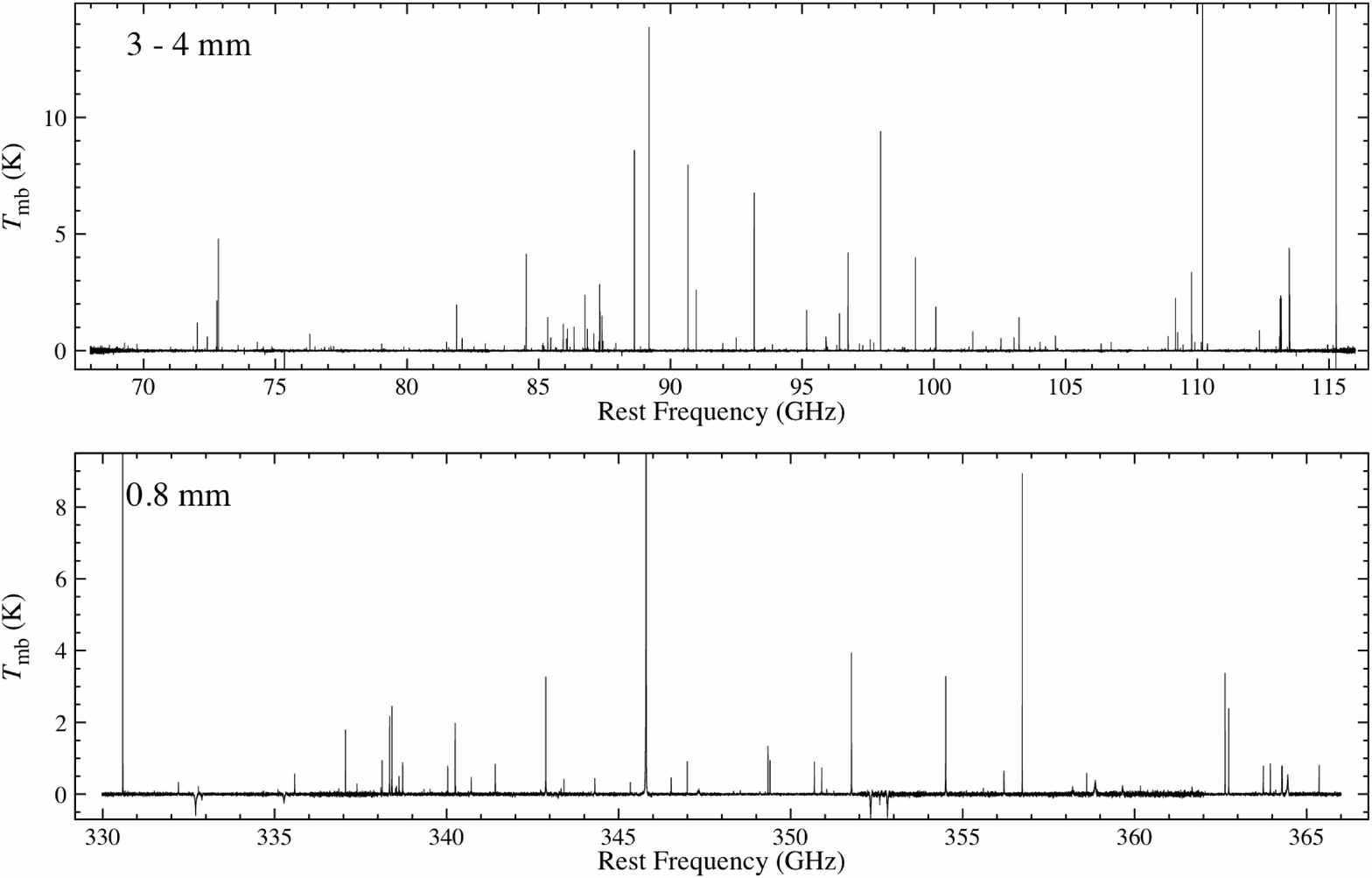}
\caption{Overall spectrum observed with the NRO 45~m telescope (upper spectrum) and that with the ASTE 10~m telescope (lower spectrum).  $V_{\rm LSR} = 7.0$ is assumed as the system velocity. }
\label{fig1}
\end{figure}

Figure~\ref{fig1} shows the overall spectra in the 4, 3, and 0.8~mm bands.  Figures~\ref{figa1} and \ref{figa2} show their expanded spectra.  In the 3~mm and 4~mm bands, 265 emission lines are detected, while 74 emission lines are detected in the 0.8~mm band.  The line detection criterion is that  a peak intensity of a line exceeds three times the r.m.s. noise level at the expected frequency.

\begin{figure}
\epsscale{0.95}
\plotone{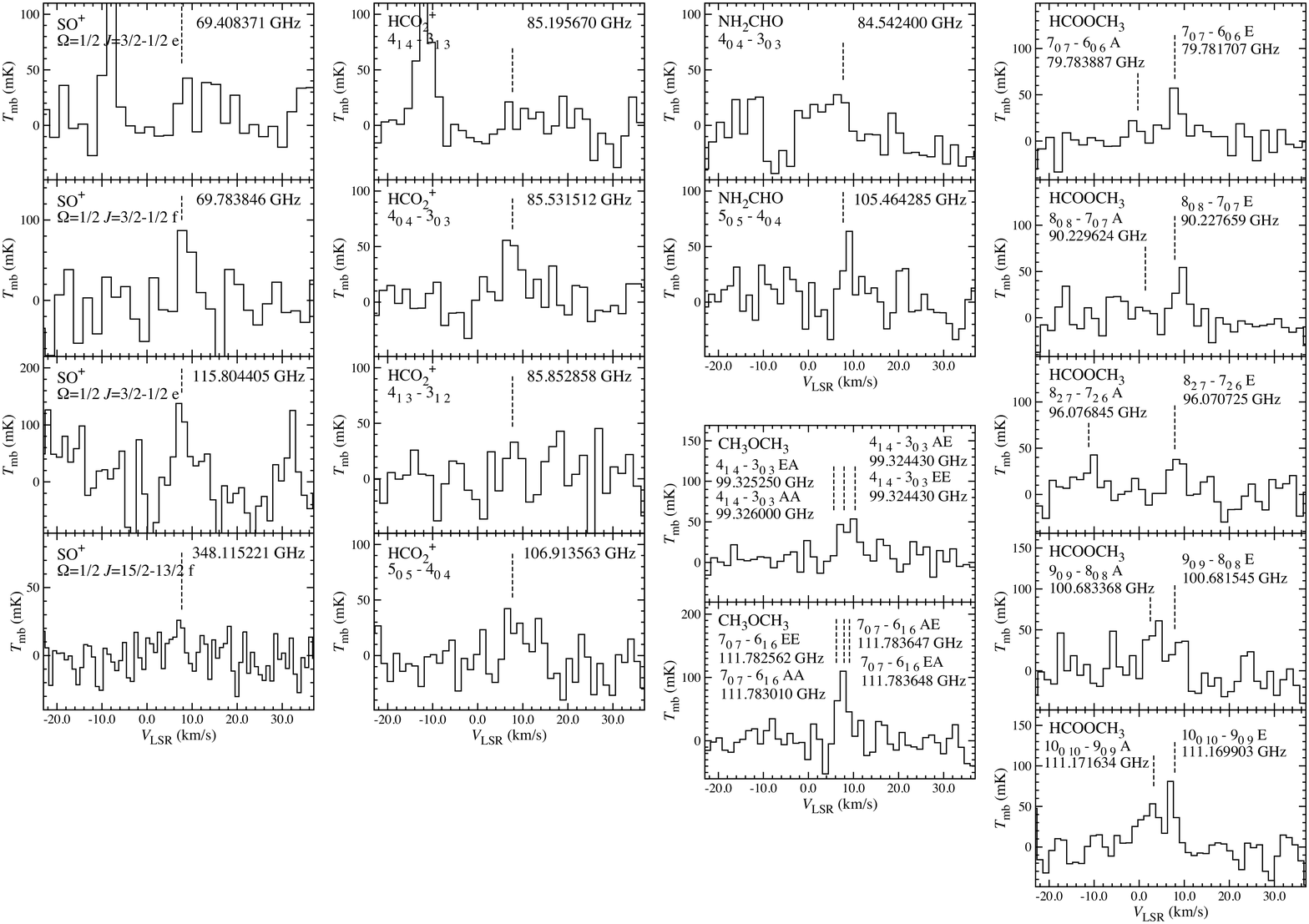}
\caption{Marginal detections of SO$^+$, HCO$_2^+$, NH$_2$CHO, CH$_3$OCH$_3$, and HCOOCH$_3$.  The vertical dotted lines indicate the expected line positions, where the line of sight velocity is 7.7~km~s$^{-1}$, according to the C$^{18}$O($J=1-0$) data. }
\label{fig2}
\end{figure}

The line identifications are carried out on the basis of the spectral line databases the Cologne Database for Molecular Spectroscopy (CDMS) managed by University of Cologne \citep{muller01,muller05} and the Submillimeter, Millimeter, and Microwave Spectral Line Catalog provided by Jet Propulsion Laboratory \citep{pickett98}.  From the detected emission lines, 35 molecular species and 29 isotopologues are identified in the 3~mm and 4~mm bands (Tabel~\ref{tab1}).  In the 0.8~mm band, 17 molecular species and 13 isotopologues are identified (Tabel~\ref{tab2}).  In total, 36 molecular species and 30 isotopologues are identified in this spectral line survey.  For the weak emission lines, we carefully check the presence of other emission lines of the same species at other frequencies. SO$^+$, NH$_2$CHO, HCO$_2^+$, HCOOCH$_3$, and CH$_3$OCH$_3$ are tentative detections, because their lines are marginally detected (Figure \ref{fig2}).  HCOOCH$_3$ was identified by \citet{Sakai2007A} in this source.  \citet{Sakai2007A} detected the HCOOCH$_3$ ($8_{1\,8}-7_{1\,7}$ A and E) lines with a peak intensity of $84 \pm 15$~mK.  Although these lines were not detected in the present survey probably due to lower sensitivity, a few other lines were marginally detected (Figure \ref{fig2}).  Several HCOOCH$_3$ lines can be seen in the E state, but most of companion HCOOCH$_3$ lines in the A state whose upper state energy and S$\mu^2$ are similar to those of the HCOOCH$_3$ in the E state are very weak.  Hence, detection of HCOOCH$_3$ is tentative in this survey.  As for CH$_3$OCH$_3$, two weak lines were marginally detected in this survey, since four sub-states of internal rotation are almost overlapped at the frequencies.  On the other hand, other transition lines of CH$_3$OCH$_3$ are not detected due to lower sensitivity and detection of CH$_3$OCH$_3$ is also tentative.  The detected molecules including tentatively detected ones are summarized in the Table \ref{tabmol}.  In spite of the above identification process, 22 lines are still unidentified in the 4 and 3~mm bands.  They are also listed in Table~\ref{tab1}.  The criterion of the unidentified line is that the line is detected with $3 \sigma$ confidence or higher and the line width is reasonable.  Since the criterion is not stringent, some of them may be spurious  signals.  Hence, the data of the unidentified lines should be used carefully.  On the other hand, no unidentified lines are found in the 0.8~mm band.   The line-of-sight velocity and FWHM line width of each line are evaluated by a single Gaussian fit.  If the line profile is blended with multiple lines such as nearby hyperfine components, multiple Gaussian functions are employed to determine the line parameters.  Tables \ref{tab1} and \ref{tab2} present lists of the line parameters of the identified molecular lines including tentatively detected ones.  

In the 3~mm and 4~mm bands, we detected many lines of various carbon-chain molecules, which include C$_4$H, HC$_3$N, HC$_5$N, CCS, and C$_3$S.  In general, carbon-chain molecules are less abundant in star-forming regions than in young starless cores \citep[e.g.][]{Suzuki1992, Hirota2009}. This is particularly true for long carbon-chain molecules.  Hence, detection of various carbon-chain molecules in NGC~2264~CMM3 should be noteworthy.  In addition to carbon-chain molecules, saturated molecules such as CH$_3$OH and CH$_3$CHO are also detected.  However, complex organic molecules such as HCOOCH$_3$ and CH$_3$OCH$_3$ seem to be deficient in this source, since these molecules are marginally detected.  Furthermore, nitrogen bearing complex organic molecules such as C$_2$H$_3$CN and C$_2$H$_5$CN were not detected in the present line survey.  

Another characteristic feature of the NGC~2264~CMM3 spectrum is the relatively bright emission of deuterated species such as DCO$^+$, DCN, DNC, and N$_2$D$^+$ in the 4~mm band.  Even the spectral lines of CCD and DC$_3$N are weakly detected.  Moreover, five lines of NH$_2$D are seen in the 0.8~mm, 3~mm, and 4~mm bands.

In the 0.8~mm band, almost all detected emission lines are higher rotational transition lines of molecular species detected in the 4~mm and 3~mm bands, except for NO (nitric oxide).  SO$_2$ is the heaviest molecule observed in this band.  No heavy molecules, which consist of more than 3 heavy atoms, were detected except for c-C$_3$H$_2$ and SO$_2$ in the 0.8~mm band.  For larger molecules, rotational spectral lines in the 0.8~mm band generally have higher upper-state energies, and they are not well excited except for the vicinity of the protostar.  Hence, these lines are usually weak in the single-dish observation in the 0.8~mm band in ordinary star-forming regions \citep[e.g.][]{Watanabe2012}.  On the other hand, higher excitation lines of CH$_3$OH and H$_2$CO, whose upper state energies ($E_{\rm u}$) are higher than 150~K, were detected.  These molecules would be abundant in a hot and dense region in the vicinity of a protostar and/or in shocked regions caused by outflows, indicating a sign of a hot core.

\subsection{Rotation Temperatures and Column Densities} 
To investigate molecular abundances and physical conditions of the emitting region, we derive rotation temperatures and column densities for molecules for which multiple transition lines with different upper-state energies are detected.  The rotation temperature and the column density are estimated under local thermodynamic equilibrium (LTE) conditions by using a least-squares method with the following formula:
\begin{equation}
\Delta T = \frac{h\nu}{k}\left[\frac{1}{\exp(h\nu / kT_{\rm rot}) - 1} - \frac{1}{\exp(h\nu / kT_{\rm bg}) - 1}\right][1-\exp(-\tau)],
\label{eq1}
\end{equation}
and
\begin{equation}
\tau = \frac{8 \pi^3 S\mu^2 N}{3k\Delta v Q(T_{\rm rot})}\left[ \exp\left(\frac{h\nu}{kT_{\rm rot}}\right) -1 \right]\exp\left(-\frac{E_{\rm u}}{kT_{\rm rot}}\right),
\label{eq2}
\end{equation}
where $\Delta T$, $k$, $\nu$, $h$, $T_{\rm rot}$, $T_{\rm bg}$, $S$, $\mu$, $N$, $\Delta v$, $Q(T)$, and $E_{\rm u}$ are line intensity, the Boltzmann constant, transition frequency, the Planck constant, rotation temperature, the cosmic microwave background temperature of 2.7~K, line strength, dipole moment, total column density, line width, partition function, and upper state energy of the transition, respectively.  In order to take frequency dependence of the beam sizes into account, the observed integrated intensities are divided by the beam filling factor $\theta_{\rm source}^2/(\theta_{\rm beam}^2+\theta_{\rm source}^2)$, where $\theta_{\rm source}$ and $\theta_{\rm beam}$ are the assumed source size and the FWHM beamwidth of the telescopes, respectively.  Here, the source size ($\theta_{\rm source}$) is assumed to be $15''$, which is the smallest beam size of this survey.  This source size is almost comparable to the size of the distribution of HCOOCH$_3$ \citep{Sakai2007A}.  The beam size $\theta_{\rm beam}$ is evaluated at each transition frequency by extrapolating the beam size measured at some representative frequencies.  The common line width ($\Delta v$) is assumed to be  3~km~s$^{-1}$ except for SiO ($\Delta v = 6$~km~s$^{-1}$), and $\int T_{\rm mb} dv/\Delta v$ is used as the line intensity of $\Delta T$.  The error of $\Delta T$ includes the r.m.s. noise and the 20~\% uncertainty of the intensity calibration by chopper-wheel method.  For NH$_2$D, H$_2$CO, c-C$_3$H$_2$, H$_2$CCO, and H$_2$CS, ortho and para species are analysed separately.  For CH$_3$CCH and CH$_3$CN, A and E states are analysed separately.  Although multiple transition lines of HC$_3$N and C$_3$S are detected, we failed to derive their rotation temperatures and column densities of these molecules with this method.  For HC$_3$N, the rotation temperature and the column density could not be determined simultaneously by the least-square fit, probably because of the low excitation temperature and relatively high optical depth ($\tau \sim 1$).  For C$_3$S, the signal-to-noise ratio is not enough to determine the excitation temperature and the column density simultaneously.  Hence, column densities of these two molecules are estimated with fixed excitation temperatures in the latter section.  Table~\ref{tab3} shows the results of the analyses.  

The rotation temperatures are found to be different from molecule to molecule, ranging from 7~K to 122~K.  Carbon-chain molecules and fundamental species such as H$^{13}$CN, HC$^{18}$O$^+$, and CN show relatively low rotation temperatures ($<11$~K).  HC$_5$N shows relatively higher rotation temperature of 25.8~K.  Because the upper state energies of the HC$_5$N lines observed in this survey are from 44~K to 110~K, which are higher than those of the other carbon-chain molecules, the lines would preferentially trace warmer region.  Deuterated species, DCN, DCO$^+$ and NH$_2$D (para), show a similar trend of low rotation temperature.  On the other hand, sulfur-bearing molecules except for sulfur-bearing carbon-chain molecule (CCS) show higher rotation temperatures ($15 < T < 26$~K).  The rotation temperature of SiO is also as high as 18.2~K.  This means that the sulfur-bearing molecules and SiO preferentially reside in a warmer and denser part than the other molecular species.  More importantly, we found that the CH$_3$OH lines cannot be fitted by a single temperature, and hence, we employed the two-component model for CH$_3$OH with the rotation temperatures of 24~K and 122~K, as discussed later in detail.  H$_2$CO also shows the relatively high rotation temperature ($33\pm13$~K) in spite of its large uncertainty.  It should be noted that the o/p ratio of NH$_2$D, H$_2$CS, H$_2$CCO, and c-C$_3$H$_2$ are close to the statistical value of 3.  

For the other molecules for which only one transition line or hyperfine lines with almost the same upper-state energies are detected, the column densities are estimated under the LTE condition with excitation temperatures of 10~K, 15~K, and 20~K by using the least-squares method with equations (\ref{eq1}) and (\ref{eq2}).  The range of the assumed excitation temperature is set on the basis of the rotation temperatures of various molecules (Table \ref{tab3}).  In order to correct beam dilution effect, the source size of $15''$ is also assumed as in the case of the rotation diagram analysis.  Uncertainties of the derived column densities include the r.m.s. noise and the intensity calibration uncertainty of 20~\%.  Table~\ref{tab4} summarizes the column densities obtained in the LTE analysis.  

The gas kinetic temperature can be estimated by using the observed intensities of the different $K$ lines of CH$_3$CN, because the radiation processes between the different $K$ ladders are almost forbidden.  The gas kinetic temperature thus obtained from the $K$ structure lines is $37 \pm 10$~K and $25 \pm 10$~K by using the $K=0$ and 3 lines (A-state) and the $K=1$ and 2 lines (E-state), respectively.  Similarly, the excitation temperature between the different $K_{\rm a}$ levels of H$_2$CO is close to the gas kinetic temperature.  It is as high as $66\pm 14$~K and $51 \pm 13$~K for the $K_{\rm a}=1$ and 3 (ortho) and $K_{\rm a}=0$, 2 and 4 (para) lines, respectively.  This result may further suggest that H$_2$CO mainly resides in a higher temperature component in the vicinity of the protostar than CH$_3$CN.  This result indicates that NGC2264~CMM3 involves physical and chemical complexity within the beam sizes of the present observations.  

\section{Discussion} 
\subsection{High Excitation Lines of CH$_3$OH} 
\begin{figure}
\epsscale{0.95}
\plotone{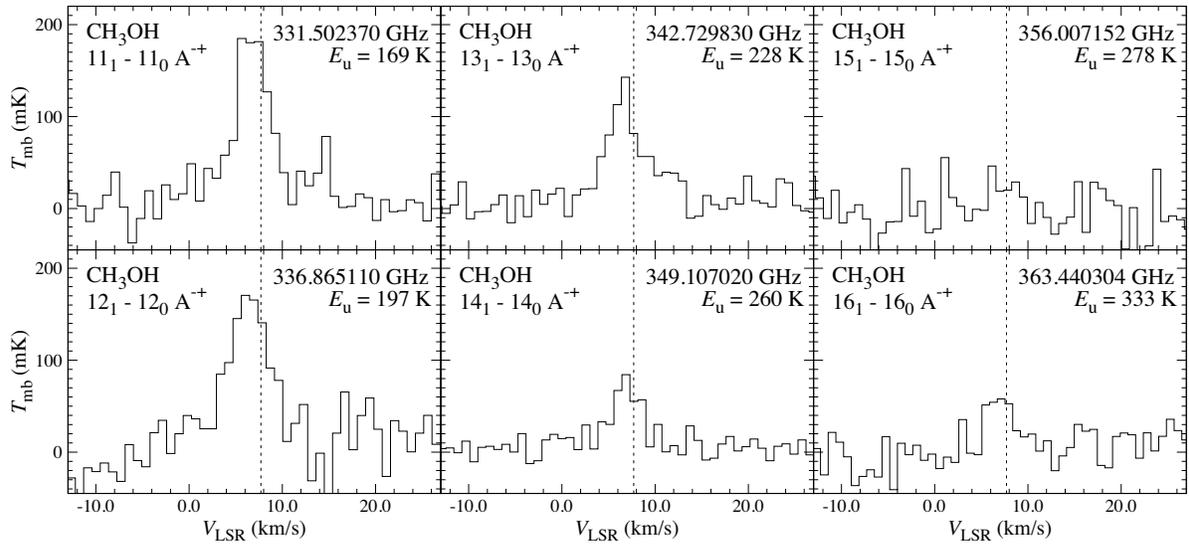}
\caption{Profile of the high excitation CH$_3$OH lines.  The vertical line indicates 7.7~km~s$^{-1}$ which is the line of sight velocity of C$^{18}$O($J=1-0$). }
\label{fig4}
\end{figure}

\begin{figure}
\epsscale{0.50}
\plotone{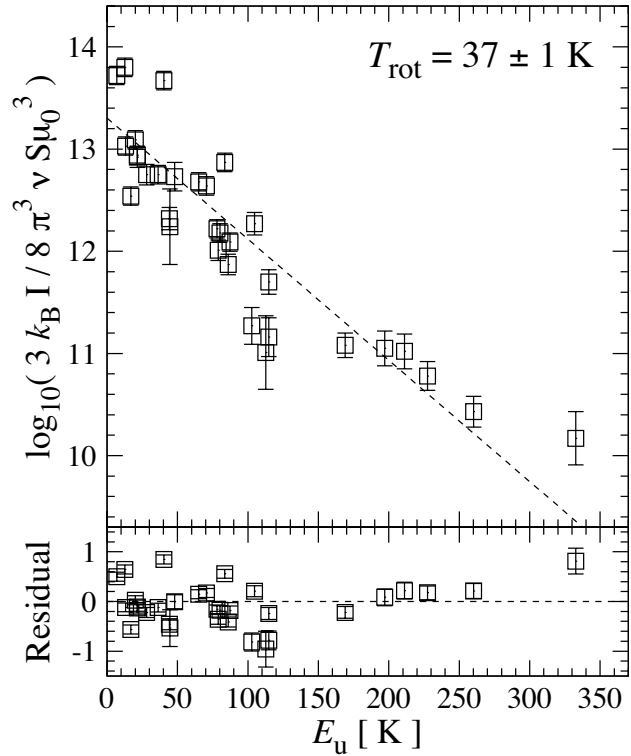}
\caption{A rotation diagram of CH$_3$OH.  A dashed line indicates a single temperature fit.  Since the single temperature fit shows a `V-shaped' systematic residual as shown in the bottom panel, we employ a two-component model in derivation of the column density, as described in the text. }
\label{fig5}
\end{figure}

In the 0.8~mm band, high excitation lines of CH$_3$OH with the upper-state energies higher ($E_{\rm u}$) than 150~K were detected (Figure \ref{fig4}).  The upper-state energies of all the other molecular lines detected in this survey are lower than 150~K except for several lines of H$_2$CO.  The rotation diagram with a single temperature model shows systematic residuals indicating coexistence of cold and warm components (Figure~\ref{fig5}).  Such a behavior in the rotation diagram is sometimes found in hot core regions.  \citet{Bisschop2007} suggested that the two components can appear from sub-thermal excitation and optical depth effects.  However, the optical depths estimated from the column density and rotation temperature by the rotation diagram method with a single component are 0.2--0.1 for transitions with the upper-state energies of 90--150~K, where the rotation diagram shows a knee structure.  The optical depths are found to be higher for the transitions with $E_{\rm u}<60$~K.  Therefore, the optical depth effect is not likely the case for NGC~2264~CMM3 as the origin of the systemic residual.  Hence, we employ a two-temperature model by extending Eq. (\ref{eq1}) for the two-components, and the rotation temperatures of the two components are derived by the least-squares fit on all the observed lines of CH$_3$OH.  The rotation temperature are derived to be $24.3 \pm 2.6$~K and $122 \pm 63$~K.  The higher temperature component seems to correspond to the hot component in the vicinity of the protostar, although the error of the temperature is large.  

%

\begin{figure}
\epsscale{0.95}
\plotone{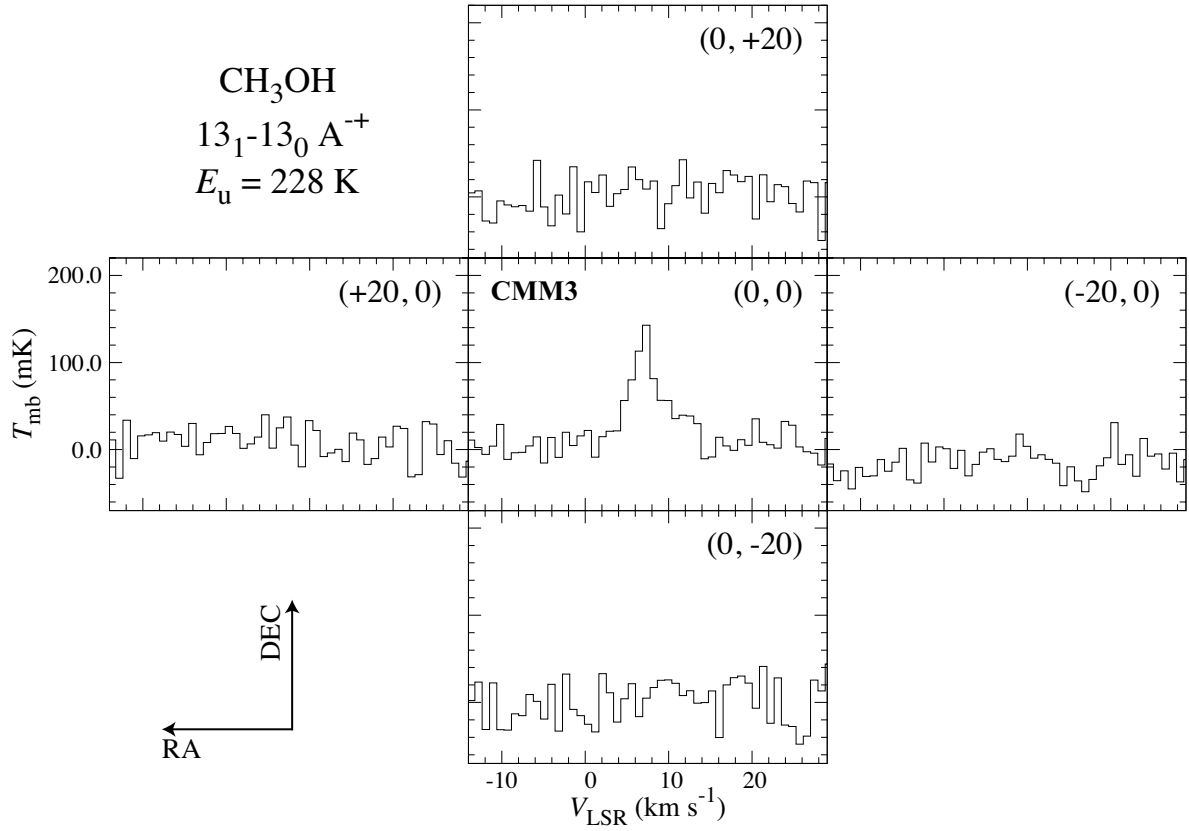}
\caption{A profile map of CH$_3$OH($13_{1}-13_{0}\,{\rm A}^{-+}$) in NGC~2264~CMM3.  The grid spacing almost corresponds to the FWHM size of the telescope beam ($22''$).  Arrows in the bottom left corner indicate directions of the right ascension (RA) and the declination (DEC).}
\label{fig6}
\end{figure}

Figure~\ref{fig6} shows a profile map of CH$_3$OH ($13_{1}-13_{0}\,{\rm A}^{-+}$) around NGC~2264~CMM3.  The distribution of high excitation lines of the CH$_3$OH is concentrated at the CMM3 position.  Therefore, it is confirmed that these CH$_3$OH lines originate from hot molecular gas in the vicinity of the protostar CMM3.  These CH$_3$OH lines are often detected in hot cores \citep[e.g.][]{vandertak2000} and hot corinos \citep[e.g.][]{Maret2005}.  In addition to heating by radiation from the protostar, shock heating may also contribute, since the high excitation lines of CH$_3$OH are also detected in the shocked regions induced by the outflow driven by the protostars in L1448-mm/IRS~3 \citep{Jimenez2004} and L1157 \citep{Codella2010}.  \citet{Saruwatari2011} reported the detection of CH$_3$OH ($5_0-4_0, {\rm A^+}$ and $5_3-4_3, {\rm A^+}$) associated with the compact outflow from the protostar in this source.  From these results, the detection of high excitation CH$_3$OH lines represents the existence of the high temperature molecular gas affected by protostellar activities of CMM3.  This is consistent with the high gas kinetic temperature inferred from the excitation temperatures between the different $K_{\rm a}$ ladders of H$_2$CO.  

\subsection{Deuterium Fractionation Ratios} 
Seven deuterated molecular species, DCO$^+$, N$_2$D$^+$, DCN, DNC, CCD, HDCO, and DC$_3$N, are identified in this line survey.  Ground-state transition lines in the 70~GHz band are observed for DCO$^+$, N$_2$D$^+$, DCN, DNC, and DCO$^+$.  Higher transition lines of DCO$^+$, DCN, and DNC are also identified in the 0.8~mm band.  HDCO is detected only in the 0.8~mm band.  Deuterated CH$_3$OH and multiply deuterated molecular species such as D$_2$CO and CHD$_2$OH are not detected, although these molecules are often found in the hot corino sources and hot core sources \citep[e.g.][]{Mauersberger1988,Turner1990,Ceccarelli1998,Parise2006}.  This non-detection would be due to insufficient sensitivity of the present line survey.  For example, the intensity of CH$_2$DOH ($2_{0\,2}-1_{0\,1}\,{\rm e}_1$) is estimated to be 2~mK in $T_{\rm mb}$ with the rotation temperature of 30~K under the LTE approximation, assuming the CH$_2$DOH/CH$_3$OH ratio of 1~\% ($2.1 \times 10^{13}$~cm$^{-2}$).  It is well below our sensitivity (r.m.s $\sim 30$~mK at the corresponding frequency).

Table~\ref{tabd} shows the deuterium fractionation ratios evaluated from the column densities derived in section~3.2.  The ratios range from 0.01 to 0.04, and no systematic trend can be seen among molecular species: the deuterium fractionation ratios of ionic molecules are similar to those of neutral molecules.  Deuterated molecules would mostly reside in a cold ambient envelope, because the rotation temperatures of DCN and NH$_2$D are estimated to be lower than 10~K (Table \ref{tab3}).  

The most characteristic feature of the deuterium fractionation ratio in this source is relatively high DCO$^+$/HCO$^+$ and N$_2$D$^+$/N$_2$H$^+$ ratios in spite of the active star-forming activities.  Deuterium fractionation ratios of the ionic species are generally low for a warm region ($T>20$~K), because the deuterium fractionation process is not efficient above 20~K and the deuterated molecular ions which had been formed in the cold stages are quickly destroyed \citep{TSakai2012,Fontani2014}.  For this reason, the DCO$^+$/HCO$^+$ and N$_2$D$^+$/N$_2$H$^+$ ratios are generally low.  Indeed, the DCO$^+$ and N$_2$D$^+$ lines are not detected in Orion~KL (Section 4.3), and the DCO$^+$/HCO$^+$ ratio is as low as 0.008 toward the low-mass star-forming region IRAS~16293-2422 \citep{vanDishoeck1995} (Table \ref{tabd}).  The moderate deuterium fractionation ratios of the ionic species mean that NGC~2264~CMM3 is surrounded by a cold envelope ($T \sim 10$~K).  In fact, the deuterium fractionation ratios of various molecules including neutral species are almost comparable with those reported for the cold dark cloud L134N \citep{Tine2000,Turner2001}, as shown in Table~\ref{tabd}.  Such a structure of NGC~2264~CMM3 seems to be related to the youth of the protostar suggested by the short dynamical age of the outflow \citep{Saruwatari2011}.  \citet{Emprechtinger2009} reported that the N$_2$D$^+$/N$_2$H$^+$ ratio decreases as protostellar evolution.  The above observational result for N$_2$D$^+$/N$_2$H$^+$ in NGC2264~CMM3 is consistent with theirs.

\subsection{Comparison with Orion KL} 
\begin{figure}
\epsscale{1.00}
\plotone{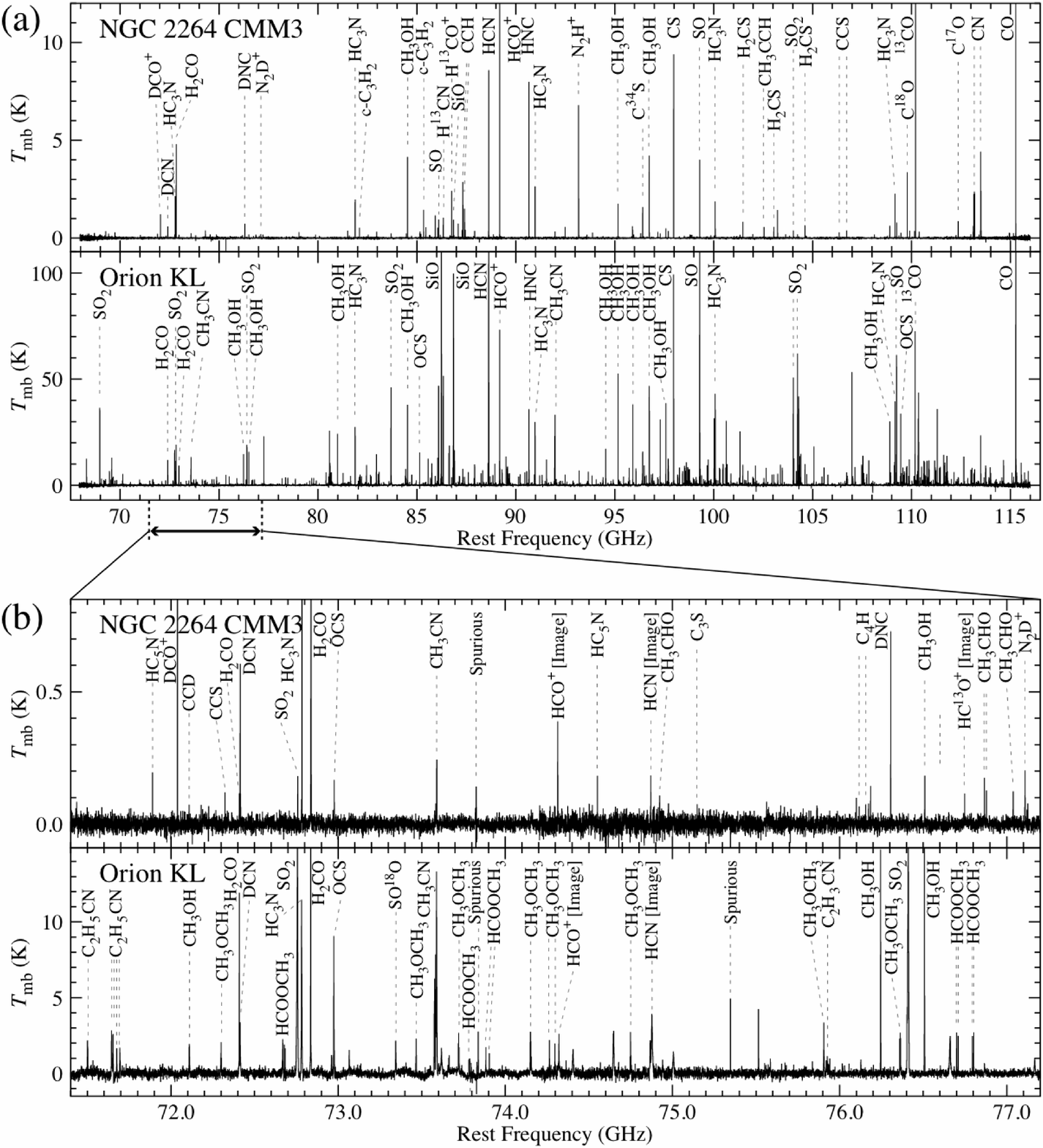}
\caption{(a) Spectra of NGC~2264~CMM3 (top) and Orion~KL (bottom) in the 4 and 3~mm bands.  (b) The expanded spectra of the two objects in the frequency range from 71.4~GHz to 77.2~GHz.  The $V_{\rm LSR}$ values are assumed to be 7.0~km~s$^{-1}$ and $5.5$~km~s$^{-1}$ for NGC~2264~CMM3 and Orion~KL, respectively.}
\label{fig7}
\end{figure}

In this study, we also observed Orion~KL in the 3~mm and 4~mm bands with the Nobeyama 45~m telescope for comparing with the NGC~2264~CMM3 spectrum.  The observed position is : ($\alpha_{\rm J2000}$, $\delta_{\rm J2000}$) = ($5^{\rm h}\,35^{\rm m}\,14^{\rm s}.5$, $-05^{\circ}\,22'\,30''.4$).   The Orion KL is the nearest high-mass star-forming region to the Sun \citep[$d=437$~pc:][]{Hirota2007}.  The chemical compositions of Orion~KL have extensively been studied with single-dish telescopes \citep[e.g.][]{Johansson1984,Turner1989,Schilke1997,Tercero2010,Tercero2011} and interferometers \citep[e.g.][]{Blake1996,Beuther2005}.  The composite spectrum is shown in Figure~\ref{fig7}.  The spectral pattern of NGC~2264~CMM3 is largely different from that of Orion~KL, indicating a significant chemical difference as well as different excitation conditions between these two sources.  Indeed, the rotation temperatures of molecules are much higher in the Orion~KL than in NGC~2264~CMM3 \citep[e.g.][]{Blake1987}.

As shown in Figure~\ref{fig7} (a), the intensities of N$_2$H$^+$ and HNC relative to the intensity of CS are brighter in the NGC~2264~CMMs than Orion~KL.  In addition, intensities of carbon-chain molecules such as CCS, C$_3$S, C$_4$H, and HC$_5$N, as well as deuterium bearing species such as DCN and DNC are found to be stronger in the NGC~2264~CMM3 (Figure~\ref{fig7} b).  Note that the DCO$^+$ and N$_2$D$^+$ lines are not detected in Orion~KL (Figure~\ref{fig7}b).  On the other hand, intensities of the SO, SO$_2$, and SiO lines are relatively stronger in Orion~KL than NGC~2264~CMM3.  Intensities of complex organic molecules such as CH$_3$CN, CH$_3$OCH$_3$, and HCOOCH$_3$ are also higher in Orion~KL.  

\begin{figure}
\epsscale{0.65}
\plotone{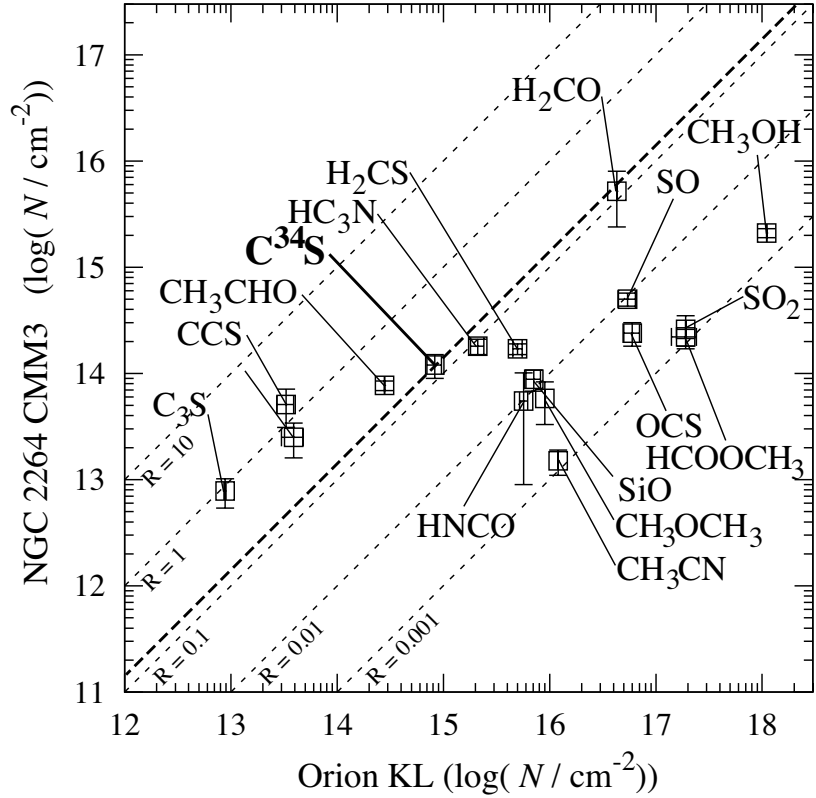}
\caption{A plot of column densities between NGC~2264~CMM3 and Orion~KL based on Table~\ref{tab5}.  For the column densities toward Orion~KL, see the footnote of Table~\ref{tab5}.  Dotted lines indicate the column density ratios of 10, 1, 0.1, 0.01 and 0.001.  The thick dashed line represents the condition that the abundances relative to C$^{34}$S are identical for the both sources.  For molecules above this line, their abundances are higher in NGC~2264~CMM3 than Orion~KL.  For the molecules below this line, their abundances are lower in NGC~2264~CMM3 than Orion~KL.  }
\label{fig8}
\end{figure}

The spectra of the two sources are strikingly different from each other, indicating the difference of chemical compositions between the two objects.  Figure~\ref{fig8} shows comparison of the column densities between NGC~2264~CMM3 and Orion~KL (Table~\ref{tab5}).  In this comparison, we employ the column densities of Orion~KL reported in the literatures \citep{Bell2014, Carvajal2009, Comito2005, Esplugues2013a, Esplugues2013b, Haykal2014, Kolesnikova2014, Marcelino2009, Neill2013, Tercero2010, Tercero2011, Turner1991}, because our observation of this source is less sensitive than the previous observations and restricted within the narrow frequency range.  Since Orion~KL is composed of several components with different physical and chemical conditions \citep{Blake1987}, we use the average column density of these component for comparison, as employed by \citet{Fuente2014}.  As described in the footnote of Table~\ref{tab5}, we calculate the column density convolved with the $15''$ beam by considering the column density and source size of each components.  This enables us a fair comparison with NGC~2264~CMM3, where several components are not resolved in this study.  

In general, the column densities are higher in Orion~KL than NGC~2246~CMM3 for almost all the species.  For comparison between the two sources, we employ C$^{34}$S as a standard molecule, because CS is ubiquitously present under various conditions \citep[e.g][]{Zhou1989, Tatematsu1993, Blake1996}.  The abundances of carbon-chain molecules such as CCS, C$_3$S, and HC$_5$N relative to the C$^{34}$S are higher by an order of magnitude in NGC~2264~CMM3 than in Orion~KL.  The abundances of SO, SO$_2$, SiO, and complex organic molecules (HCOOCH$_3$ and CH$_3$OCH$_3$) are lower in NGC~2264~CMM3 by an order of magnitude or more.  An exception for this is CH$_3$CHO, which shows little difference between the two objects.

\begin{figure}
\epsscale{0.95}
\plotone{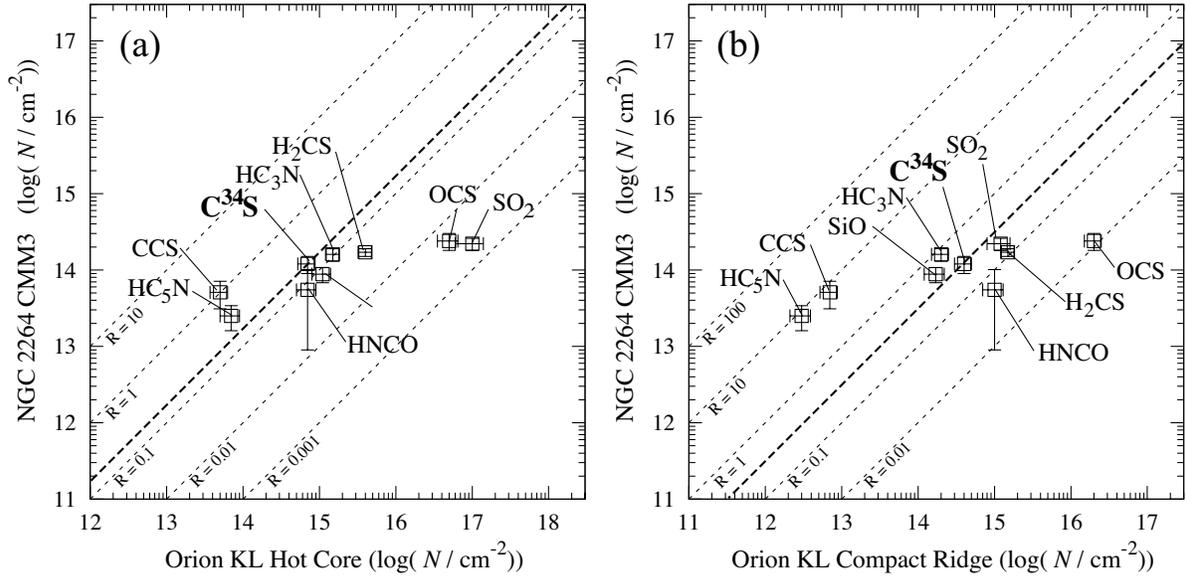}
\caption{(a) A plot of column densities between NGC~2264~CMM3 and the Orion hot core.  Dotted lines indicate the column density ratios of 10, 1, 0.1, 0.01 and 0.001.  The thick dashed line represents the condition that the abundances relative to C$^{34}$S and identical for the both sources.  (b) A plot of column densities between NGC~2264~CMM3 and the Orion compact ridge.  Dotted lines indicate the column density ratios of 100, 10, 1, 0.1, and 0.01.  The meaning of the thick dashed line is the same as (a).  The column densities of Orion hot core and compact ridge are employed from \citet{Marcelino2009} (HNCO), \citet{Tercero2010} (C$^{34}$S, H$_{2}$CS, and OCS), \citet{Tercero2011} (SiO), \citet{Esplugues2013a} (SO and SO$_2$), and \citet{Esplugues2013b} (HC$_3$N, and HC$_5$N).  }
\label{fig9}
\end{figure}

For reference, we prepared the column density plot between NGC2264~CMM3 and the Orion hot core and that between NGC2264~CMM3 and the Orion compact ridge, as shown in Figures~\ref{fig9} (a) and (b), respectively.  In this case, the averaging processes made for Figure~\ref{fig8} are not adopted.   Even in this plot, we can see the same trend mentioned above, although the number of molecular species are limited, because some species only exist in the hot core or the compact ridge.  

From the comparisons with Orion~KL, NGC~2264~CMM3 is found to be abundant in carbon-chain molecules and deficient in complex organic molecules, SO, and SO$_2$.  Abundant carbon-chain molecules are usually found in young starless cores \citep{Suzuki1992,Hirota2009}, while complex organic molecules and sulfur-bearing molecules are characteristic of hot cores and hot corions \citep[e.g.][]{Blake1994, vanDishoeck1995, Cazaux2003}.  Therefore, the chemical compositions found in NGC~2264~CMM3 indicate its chemical youth, where the protostar is very young and a hot core is not yet well developed around it.  Indeed, the protostar is deeply embedded in the dense core, as indicated by high visual extinction to the protostar \citep{Saruwatari2011}.   The chemical youth is also suggested by the richness in deuterated molecules: NGC~2264~CMM3 is still surrounded by cold envelope gas.  Although high excitation CH$_3$OH lines are detected, most of molecules would reside in the protostar envelope, because the rotation temperatures in NGC~2264~CMM3 are lower than those found in hot cores \citep[e.g.][]{Blake1987,Favre2011}.  

It is very surprising that such a cold envelope gas still exists in the central part of the active cluster forming region as a parent cloud for high-mass star formation.  Therefore, this result would give an important clue to understanding evolution of a cluster forming clump.  It is also interesting that NGC~2264~CMM3 seems to harbor a hot and dense region around the protostar, as inferred from detection of the high excitation lines of CH$_3$OH.  It is still controversial whether such a hot and dense region corresponds to a hot core or to a shocked region caused by the infant outflow found by \citet{Saruwatari2011}.  Above all, this source is a novel target to explore the early stage of high mass star formation, and high spatial resolution observations are awaited.  

\acknowledgments
The authors are grateful to the Nobeyama Radio Observatory (NRO) staff for excellent support in the observation with the 45~m telescope.  The 45~m radio telescope is operated by the NRO, a branch of the National Astronomical Observatory of Japan, National Institutes of Natural Sciences.  The authors are also grateful to the ASTE staff for excellent support.  ASTE project is driven by the NRO in collaboration with University of Chile, and Japanese institutes including University of Tokyo, Nagoya University, Osaka Prefecture University, Ibaraki University, and Hokkaido University.  Observations with ASTE were in part carried out remotely from Japan by using NTT's GEMnet2 and its partnet R\&E (Research and Education) networks, which are based on AccessNova collaboration among University of Chile, NTT Laboratories, and NAOJ.  This study is supported by a Grant-in-Aid from the Ministry of Education, Culture, Sports, Science, and Technology of Japan (No. 21224002, 21740132, and 25108005).




\clearpage

\clearpage
\begin{table}
\begin{center}
\small
\caption{Summary of observations at Nobeyama 45~m and ASTE 10~m.}
\begin{tabular}{llcccccc}
\tableline\tableline
Telescope     & Receiver & Frequency\tablenotemark{a}           & Res.\tablenotemark{b} & Tsys\tablenotemark{c} & r.m.s \tablenotemark{d} & 
 FWHM \tablenotemark{e} &   $\eta_{\rm mb}$\tablenotemark{f}\\
              &          & (GHz)                            & (MHz)                 & (K)                   & (mK) & (arcsec) & (\%) \\
\tableline
Nobeyama 45~m & T70H/V   &  67.97 --  92.43                 & 0.5 & 140 -- 380      & 13 -- 51 & 20 -- 22 & 31 -- 38 \\
Nobeyama 45~m & TZ1H/V   &  92.37 -- 116.00                 & 0.5 & 140 -- 240      & 11 -- 56 & 15 -- 20 & 24 -- 31\\
ASTE 10~m     & CATS345  & 329.98 -- 366.00                 & 1.0 & 150 -- 400      & 13 -- 39 & 21 -- 23 & 60  \\
\tableline
\end{tabular}
\label{tab0}
\tablenotetext{a}{Observed frequency ranges.}
\tablenotetext{b}{The spectral resolutions of the final spectra.}
\tablenotetext{c}{The system noise temperatures of the receivers in the observations.}
\tablenotetext{d}{A typical r.m.s. noise of the final spectra at the spectral resolution.}
\tablenotetext{e}{The full-width half-maximum beam width.}
\tablenotetext{f}{The main-beam efficiencies of the telescopes.}
\end{center}
\end{table}

\clearpage
\begin{deluxetable}{lrlrrllll}
\tabletypesize{\tiny}
\tablecolumns{9}
\tablecaption{Line parameters observed in NGC~2264~CMM3 with NRO 45~m telescope.}
\tablehead{
\colhead{Name \tablenotemark{a}} & \colhead{Frequency \tablenotemark{b}} & \colhead{Transition} & \colhead{$E_{\rm u}$} & \colhead{S$\mu^2$} & \colhead{$T_{\rm mb}$ Peak \tablenotemark{c}} & \colhead{$\int T_{\rm mb} dv$ \tablenotemark{c}} & \colhead{FWHM \tablenotemark{d}} & \colhead{$V_{\rm LSR}$ \tablenotemark{d}} \\
\colhead{} & \colhead{(GHz)} & \colhead{} & \colhead{(K)} & \colhead{(Debye$^{2}$)} & \colhead{(mK)} & \colhead{(K~km~s$^{-1}$)} & \colhead{(km~s$^{-1}$)} &\colhead{(km~s$^{-1}$)} 
}
\startdata
CH$_3$CCH       &  68.361036 & $4_2-3_2$                              & 37.1 & 3.6865 & $< 154$           & $< 0.3$         &               &               \\
CH$_3$CCH       &  68.364957 & $4_1-3_1$                              & 15.4 & 4.6087 & $222 \pm 154$     & $0.85 \pm 0.66$ & $2.9 \pm 2.7$ & $8.1 \pm 0.7$ \\
CH$_3$CCH       &  68.366264 & $4_0-3_0$                              &  8.2 & 4.9158 & $206 \pm 154$     & $1.10 \pm 0.66$ & $4.5 \pm 1.6$ & $7.2 \pm 0.6$ \\
H$_2$CS         &  68.699385 & $2_{0\,2}-1_{0\,1}$                    &  4.9 & 5.4395 & $257 \pm 141$     & $1.15 \pm 0.42$ & $3.2 \pm 0.9$ & $7.7 \pm 0.2$ \\
SO$_2$          &  68.972159 & $6_{1\,5}-6_{0\,6}$                    & 22.5 & 14.779 & $176 \pm 115$     & $0.72 \pm 0.42$ & $3.3 \pm 1.0$ & $7.2 \pm 0.3$ \\
NS              &  69.002890 & $\Omega=1/2,J=3/2-1/2,F=5/2-3/2,e$     &  3.3 & 6.5557 & $184 \pm 107$     & $0.56 \pm 0.39$ & $2.8 \pm 0.5$ & $7.6 \pm 0.3$ \\
NS              &  69.017953 & $\Omega=1/2,J=3/2-1/2,F=3/2-1/2,e$     &  3.3 & 2.4268 & $124 \pm 107$     & $0.26 \pm 0.23$ & $2.1 \pm 1.1$ & $7.7 \pm 1.0$ \\
NS              &  69.037510 & $\Omega=1/2,J=3/2-1/2,F=3/2-3/2,e$     &  3.3 & 1.9404 & $115 \pm 107$     & $0.39 \pm 0.32$ & $3.1 \pm 1.4$ & $8.2 \pm 0.5$ \\
\enddata
\tablenotetext{a}{Unidentified lines are denoted as `U'.}
\tablenotetext{b}{Rest frequency.}
\tablenotetext{c}{The errors are 3$\sigma$.  The upper limit to the integrated intensity is calculated as: $\int T_{\rm mb} dv < 3 {\rm r.m.s} \times \sqrt{\Delta V / \Delta v_{\rm res}} \Delta v_{\rm res}$, where $\Delta V$ is the assumed line width (2~km~s$^{-1}$) and $\Delta v_{\rm res}$ is the velocity resolution per channel.}
\tablenotetext{d}{FWHM and $V_{\rm LSR}$ are obtained from Gaussian fits.}
\tablenotetext{e}{Gaussian fitting is not successful due to narrow line widths.}
\tablenotetext{f}{Gaussian fitting is not successful due to poor S/N ratios or blending with other lines.}
\tablenotetext{g}{Gaussian fitting is not successful due to blending with other lines.}
\label{tab1}
\end{deluxetable}

\clearpage
\begin{deluxetable}{lrlrrrrrr}
\tabletypesize{\tiny}
\tablecolumns{9}
\tablecaption{Line parameters observed in NGC~2264~CMM3 with ASTE.}
\tablehead{
\colhead{Name} & \colhead{Frequency\tablenotemark{a}} & \colhead{Transition} & \colhead{$E_{\rm u}$} & \colhead{S$\mu^2$} & \colhead{$T_{\rm mb}$ Peak \tablenotemark{b}} & \colhead{$\int T_{\rm mb} dv$ \tablenotemark{b}} & \colhead{FWHM \tablenotemark{c}} & \colhead{$V_{\rm LSR}$ \tablenotemark{c}} \\
\colhead{} & \colhead{(GHz)} & \colhead{} & \colhead{(K)} & \colhead{(Debye$^{2}$)} & \colhead{(mK)} & \colhead{(K~km~s$^{-1}$)} & \colhead{(km~s$^{-1}$)} &\colhead{(km~s$^{-1}$)} 
}
\startdata
$^{13}$CO      & 330.587965 & $J=3-2$                              &  31.7 & 0.03653& $18981 \pm 78$ & $72.2 \pm 0.3$  & $3.6 \pm 0.1$ & $7.4 \pm 0.1$ \\
D$_{2}$CO      & 330.674348 & $6_{1\,6}-5_{1\,5}$                  &  61.1 & 31.678 & $< 78$         & $< 0.2$         &               &               \\
CH$_3$OH       & 331.502370 & $11_{1}-11_{0},{\rm A^{-+}}$         & 169.0 & 21.311 & $185 \pm 62$   & $1.07 \pm 0.22$ & $4.6 \pm 0.4$ & $7.3 \pm 0.2$ \\
SO$_2$         & 332.505242 & $4_{3\,1}-3_{2\,2}$                  &  31.3 & 6.9211 & $64 \pm 61$    & $0.14 \pm 0.12$ & $^{\rm d}$    &               \\
NH$_2$D        & 332.781890 & $1_{0\,1},0a-0_{0\,0},0a$            &  16.6 & 0.1640 & $224 \pm 67$   & $0.65 \pm 0.17$ & $2.8 \pm 0.3$ & $7.2 \pm 0.1$ \\
NH$_2$D        & 332.822510 & $1_{0\,1},0s-0_{0\,0},0s$            &  16.0 & 0.1529 & $71 \pm 67$    & $0.30 \pm 0.15$ & $3.8 \pm 0.9$ & $7.6 \pm 0.4$ \\
SO$_2$         & 334.673353 & $8_{2\,6}-7_{1\,7}$                  &  43.2 & 4.9501 & $98 \pm 76$    & $0.14 \pm 0.12$ & $^{\rm e}$    &               \\
HDCO           & 335.096782 & $5_{1\,4}-4_{1\,3}$                  &  56.3 & 26.055 & $143 \pm 76$   & $0.37 \pm 0.17$ & $2.4 \pm 0.3$ & $6.9 \pm 0.1$ \\
\enddata
\tablenotetext{a}{Rest frequency.}
\tablenotetext{b}{The errors are 3$\sigma$.  The upper limit to the integrated intensity is calculated as: $\int T_{\rm mb} dv < 3 {\rm r.m.s} \times \sqrt{\Delta V / \Delta v_{\rm res}} \Delta v_{\rm res}$, where $\Delta V$ is the assumed line width (2~km~s$^{-1}$) and $\Delta v_{\rm res}$ is the velocity resolution per channel.}
\tablenotetext{c}{FWHM and $V_{\rm LSR}$ are obtained from Gaussian fits.}
\tablenotetext{d}{Gaussian fitting is not successful due to poor S/N ratios or blending with other lines.}
\tablenotetext{e}{Gaussian fitting is not successful due to a distorted baseline.}
\tablenotetext{f}{Gaussian fitting is not possible due to a very broad and complex line profile.}
\label{tab2}
\end{deluxetable}

\begin{table}
\begin{center}
\caption{Identified molecules.}
\begin{tabular}{l|l}
\tableline
\tableline
NH$_2$D & 
HNCO \\
CCH, CCD & 
NH$_2$CHO $^{*}$ \\
CN, $^{13}$CN, C$^{15}$N & 
HCO$_2^+$ $^{*}$ \\
HCN, H$^{13}$CN, HC$^{15}$N, DCN &
SiO \\
HNC, HN$^{13}$C, H$^{15}$NC, DNC & 
CS, $^{13}$CS, C$^{33}$S, C$^{34}$S \\
CO, $^{13}$CO, C$^{17}$O, C$^{18}$O & 
HCS$^+$ \\
HCO  &
H$_2$CS \\
HCO$^+$, H$^{13}$CO$^+$, HC$^{18}$O$^+$, DCO$^+$ & 
NS \\
H$_2$CO, H$_2^{13}$CO, HDCO  &
SO, $^{33}$SO, $^{34}$SO \\
CH$_3$OH, $^{13}$CH$_3$OH & 
SO$^+$ $^{*}$ \\
N$_2$H$^+$, $^{15}$NNH$^+$, N$_2$D$^+$ &
C$_4$H \\
NO &
HC$_3$N, H$^{13}$CCCN, HC$^{13}$CCN, HCC$^{13}$CN, DC$_3$N \\
c-C$_3$H$_2$ & 
HCOOCH$_3$ \\
CH$_3$CCH &
CCS \\
CH$_3$CN & 
OCS \\
H$_2$CCO &
SO$_2$ \\
CH$_3$CHO & 
C$_3$S \\
CH$_3$OCH$_3$ &
HC$_5$N \\ 

\end{tabular}
\label{tabmol}
\tablenotetext{*}{Tentative detection (see text).}
\end{center}
\end{table}

\begin{deluxetable}{llc|llc}
\tablecolumns{5}
\tablewidth{0pt}
\tabletypesize{\footnotesize}
\tablecaption{Column Densities and Rotation Temperatures $^{\rm a\,b}$.}
\tablehead{
\colhead{Molecule} & \colhead{$N$ (cm$^{-2}$)} & \colhead{$T_{\rm rot}$ (K)} & \colhead{Molecule} & \colhead{$N$ (cm$^{-2}$)} & \colhead{$T_{\rm rot}$ (K)}}
\startdata
NH$_2$D (para)          & $(1.4 \pm 0.6) \times 10^{14}$ & $ 9.7 \pm 1.9$ & 
c-C$_3$H$_2$ (para)     & $(3.6 \pm 1.6) \times 10^{13}$ & $ 7.2 \pm 3.0$ \\
NH$_2$D (ortho) $^{\rm c}$ & $(3.3 \pm 0.8) \times 10^{14}$ & $ 9.7 $     & 
A-CH$_3$CCH $^{\rm l}$  & $(3.3 \pm 1.7) \times 10^{14}$ & $10.5 \pm 7.0$ \\
CCH                     & $(3.5 \pm 0.4) \times 10^{15}$ & $11.2 \pm 0.7$ & 
E-CH$_3$CCH $^{\rm m}$  & $(2.9 \pm 1.9) \times 10^{14}$ & $ 9.2 \pm 5.8$ \\
CCD $^{\rm d}$          & $(1.3 \pm 1.1) \times 10^{14}$ & $11.2$         & 
A-CH$_3$CN $^{\rm l}$   & $(7.9 \pm 3.2) \times 10^{12}$ & $ 9.4 \pm 3.9$ \\
CN                      & $(1.2 \pm 0.2) \times 10^{15}$ & $9.0  \pm 0.7$ & 
E-CH$_3$CN $^{\rm m}$   & $(7.1 \pm 2.9) \times 10^{12}$ & $ 9.6 \pm 4.5$ \\
$^{13}$CN $^{\rm e}$    & $(2.7 \pm 0.6) \times 10^{13}$ & $9.0$          & 
H$_2$CCO (para)         & $(6.1 \pm 5.8) \times 10^{13}$ & $ 6.5 \pm 4.0$ \\
C$^{15}$N $^{\rm e}$    & $(9.2 \pm 3.6) \times 10^{12}$ & $9.0$          & 
H$_2$CCO (ortho) $^{\rm n}$ & $(2.2 \pm 0.9) \times 10^{13}$ & $ 6.5 $        \\
H$^{13}$CN              & $(5.8 \pm 1.1) \times 10^{13}$ & $ 9.1 \pm 0.8$ & 
CH$_3$CHO               & $(7.7 \pm 0.8) \times 10^{13}$ & $11.2 \pm 2.4$ \\
\enddata
\tablenotetext{a}{Obtained by a least-squares fit for multiple transitions.  The error denotes 3~$\sigma$.}
\tablenotetext{b}{Assumed source size and line width are $15''$ and 3~km~s$^{-1}$, respectively. }
\tablenotetext{c}{The rotation temperature is assumed to be the same as that of NH$_{2}$D (Para). }
\tablenotetext{d}{The rotation temperature is assumed to be the same as that of CCH. }
\tablenotetext{e}{The rotation temperature is assumed to be the same as that of CN. }
\tablenotetext{f}{The rotation temperature is assumed to be the same as that of H$^{13}$CN. }
\tablenotetext{g}{The rotation temperature is assumed to be the same as that of HN$^{13}$C. }
\tablenotetext{h}{The rotation temperature is assumed to be the same as that of C$^{17}$O. }
\tablenotetext{i}{The rotation temperature is assumed to be the same as that of HC$^{18}$O$^+$. }
\tablenotetext{j}{The rotation temperature is assumed to be the same as that of H$_2$CO (ortho). }
\tablenotetext{k}{The column densities and rotation temperatures are estimated with two component model. `Cold' and `Hot' indicate results for the low temperature component and the high temperature component, respectively. }
\tablenotetext{l}{The column density and rotation temperature of A state are estimated from $K=0$ lines. }
\tablenotetext{m}{The column density and rotation temperature of E state are estimated from $K=1$ lines. }
\tablenotetext{n}{The rotation temperature is assumed to be the same as that of H$_2$CCO (ortho). }
\tablenotetext{o}{The line width is assumed to be 6~km~s$^{-1}$. }
\tablenotetext{p}{The rotation temperature is assumed to be the same as that of C$^{34}$S. }
\tablenotetext{q}{The rotation temperature is assumed to be the same as that of SO. }
\label{tab3}
\end{deluxetable}

\begin{deluxetable}{lllllll}
\tablecolumns{4}
\tablewidth{0pt}
\tabletypesize{\scriptsize}
\tablecaption{Column Densities of Identified Molecules.$^{\rm a}$}
\tablehead{
\colhead{Molecule} & \colhead{$N$ ($T$=10~K)$^{\rm b}$} & \colhead{$N$ ($T$=15~K)$^{\rm b}$} & \colhead{$N$ ($T$=20~K)$^{\rm b}$} \\
\colhead{} & \colhead{(cm$^{-2}$)$^{\rm c}$} & \colhead{(cm$^{-2}$)$^{\rm c}$} & \colhead{(cm$^{-2}$)$^{\rm c}$} }
\startdata
HCO               & $(3.8 \pm 1.6) \times 10^{13}$ & $(5.0 \pm 2.1) \times 10^{13}$ & $(6.8 \pm 2.8) \times 10^{13}$\\
N$_2$H$^+$        & $(4.3 \pm 2.5) \times 10^{14}$ & $(3.6 \pm 0.9) \times 10^{14}$ & $(3.8 \pm 0.8) \times 10^{14}$\\
$^{15}$NNH$^+$    & $(1.4 \pm 0.8) \times 10^{12}$ & $(1.7 \pm 1.0) \times 10^{12}$ & $(2.0 \pm 1.2) \times 10^{12}$\\
N$^{15}$NH$^+$    & $< 3.6 \times 10^{11}$ & $< 4.3 \times 10^{11}$ & $< 5.1 \times 10^{11}$\\
N$_2$D$^+$        & $(7.8 \pm 2.2) \times 10^{12}$ & $(9.2 \pm 2.5) \times 10^{12}$ & $(1.1 \pm 0.3) \times 10^{13}$\\
NO                & $(5.5 \pm 1.6) \times 10^{15}$ & $(2.3 \pm 0.6) \times 10^{15}$ & $(1.6 \pm 0.4) \times 10^{15}$\\
CH$_3$OCH$_3$     & $(5.2 \pm 2.4) \times 10^{13}$ & $(5.8 \pm 2.5) \times 10^{13}$ & $(6.6 \pm 2.8) \times 10^{13}$\\
NH$_2$CHO         & $2.5^{+2.6}_{-2.5} \times 10^{12}$ & $2.6^{+2.7}_{-2.6} \times 10^{12}$ & $3.0^{+3.1}_{-3.0} \times 10^{12}$\\
\enddata
\tablenotetext{a}{Errors of the column densities are estimated by taking into account the r.m.s. noise and calibration uncertainties of the chopper-wheel method (20~\%).  The source size is assumed to be $15''$.}
\tablenotetext{b}{Assumed excitation temperatures.  }
\tablenotetext{c}{The upper limit to the column density is estimated from the 3$\sigma$ upper limit of the integrated intensity assuming the line width of 2~km/s.}
\label{tab4}
\end{deluxetable}

\begin{table}
\begin{center}
\scriptsize
\caption{Deuterium fractionation ratios.}
\begin{tabular}{lcccl}
\tableline\tableline
Molecule     & NGC~2264~CMM3 & L134~N     & IRAS~16293 \\
\tableline
DCO$^+$/HCO$^+$       & $0.030 \pm 0.015$ \tablenotemark{a}   & 0.18 \tablenotemark{f}              & 0.0086 \tablenotemark{h} \\
N$_2$D$^+$/N$_2$H$^+$ & $0.026 \pm 0.010$ \tablenotemark{b}   & 0.35 \tablenotemark{f}              &                   \\
DCN/HCN               & $0.010 \pm 0.003$ \tablenotemark{c}   & $0.050 \pm 0.016$ \tablenotemark{g} & 0.013 \tablenotemark{h}   \\
DNC/HNC               & $0.017 \pm 0.012$ \tablenotemark{c}   & $0.047 \pm 0.020$ \tablenotemark{g} & 0.03 \tablenotemark{h}    \\
CCD/CCH               & $0.037 \pm 0.032$                     & $0.060 \pm 0.040$ \tablenotemark{g} & 0.18 \tablenotemark{h} \\
DC$_3$N/HC$_3$N       & $0.016 \pm 0.013$ \tablenotemark{b}   &                                     &                   \\
HDCO/H$_2$CO          & $0.0020 \pm 0.0015$ \tablenotemark{d} & 0.068 \tablenotemark{f}             & 0.14 \tablenotemark{h} \\
D$_2$CO/H$_2$CO       & $< 0.0003$ \tablenotemark{de}         &                                     & $0.05 \pm 0.025$ \tablenotemark{i} \\
\tableline
\end{tabular}
\label{tabd}
\tablenotetext{a}{The column density of the normal molecule is estimated from the $^{18}$O isotopologue assuming that $^{16}$O/$^{18}$O = 560.}
\tablenotetext{b}{The column density is estimated under the LTE approximation with $T_{\rm ex}$ of 15~K.}
\tablenotetext{c}{The column density of the normal molecule is estimated from the $^{13}$C isotopologue assuming that $^{12}$C/$^{13}$C = 60.}
\tablenotetext{d}{The column density of the normal molecule is estimated from the H$_2^{13}$CO (para) data assuming that the ortho to para is 3 and the $^{12}$C/$^{13}$C ratio of 60.}
\tablenotetext{e}{The column density of the D$_2$CO is estimated from the D$_2$CO (ortho) data assuming that the ortho to para ratio is 2.}
\tablenotetext{f} {\citet{Tine2000} }
\tablenotetext{g} {\citet{Turner2001} }
\tablenotetext{h} {\citet{vanDishoeck1995} }
\tablenotetext{i} {\citet{Parise2006} }
\end{center}
\end{table}

\begin{deluxetable}{llll}
\tablecolumns{4}
\tablewidth{0pt}
\tabletypesize{\footnotesize}
\tablecaption{Column Densities Observed in NGC~2264~CMM3 and Orion~KL.}
\tablehead{
\colhead{Molecule} & \colhead{NGC~2264~CMM3  } & \colhead{Orion~KL$^{\rm a}$ } & \colhead{References $^{\rm b}$}\\
\colhead{}         & \colhead{$N$ (cm$^{-2}$)} & \colhead{$N$ (cm$^{-2}$)} & \colhead{} 
}
\startdata
H$_2$CO       & $(5.2 \pm 2.8) \times 10^{15}$            & $4.3 \times 10^{16}$ $^{\rm gh}$          & \citet{Neill2013} \\
CH$_3$OH      & $(2.1 \pm 0.2) \times 10^{15}$ $^{\rm c}$ & $1.1 \times 10^{18}$ $^{\rm gh}$          & \citet{Kolesnikova2014} \\
CH$_3$CN      & $(1.5 \pm 0.4) \times 10^{13}$            & $1.2 \times 10^{16}$ $^{\rm gh}$          & \citet{Bell2014} \\
CH$_3$CHO     & $(7.7 \pm 0.8) \times 10^{13}$            & $2.8 \times 10^{14}$ $^{\rm g}$           & \citet{Turner1991} \\
CH$_3$OCH$_3$ & $(5.8 \pm 2.5) \times 10^{13}$ $^{\rm d}$ & $9.0 \times 10^{15}$ $^{\rm g}$           & \citet{Comito2005} \\
HNCO          & $(5.5 \pm 4.6) \times 10^{13}$            & $(5.7 \pm 0.5) \times 10^{15}$ $^{\rm e}$ & \citet{Marcelino2009} \\
SiO           & $(8.8 \pm 2.0) \times 10^{13}$            & $(7.1 \pm 1.1) \times 10^{15}$ $^{\rm e}$ & \citet{Tercero2011}  \\
C$^{34}$S     & $(1.2 \pm 0.3) \times 10^{14}$            & $(8.3 \pm 1.3) \times 10^{14}$ $^{\rm e}$ & \citet{Tercero2010} \\
H$_2$CS       & $(1.7 \pm 0.2) \times 10^{14}$            & $(5.0 \pm 0.5) \times 10^{15}$ $^{\rm e}$ & \citet{Tercero2010}  \\
SO            & $(5.0 \pm 0.7) \times 10^{14}$            & $(5.4 \pm 1.0) \times 10^{16}$ $^{\rm e}$ & \citet{Esplugues2013a}  \\
HC$_3$N       & $(1.6 \pm 0.3) \times 10^{14}$ $^{\rm d}$ & $(2.1 \pm 0.3) \times 10^{15}$ $^{\rm f}$ & \citet{Esplugues2013b} \\
HCOOCH$_3$    & $(2.6 \pm 0.9) \times 10^{14}$ $^{\rm d}$ & $2.9 \times 10^{17}$ $^{\rm fgh}$         & \cite{Carvajal2009}, \citet{Haykal2014} \\
CCS           & $(5.1 \pm 2.0) \times 10^{13}$            & $(3.3 \pm 0.6) \times 10^{13}$ $^{\rm e}$ & \citet{Tercero2010}  \\
OCS           & $(2.4 \pm 0.6) \times 10^{14}$            & $(6.0 \pm 1.0) \times 10^{16}$ $^{\rm e}$ & \citet{Tercero2010}  \\
SO$_2$        & $(2.2 \pm 0.4) \times 10^{14}$            & $(1.9 \pm 0.5) \times 10^{17}$ $^{\rm e}$ & \citet{Esplugues2013a} \\
C$_3$S        & $(7.8 \pm 2.4) \times 10^{12}$ $^{\rm d}$ & $(8.8 \pm 0.2) \times 10^{12}$ $^{\rm e}$ & \citet{Tercero2010}  \\
HC$_5$N       & $(2.5 \pm 0.9) \times 10^{13}$            & $(3.9 \pm 0.9) \times 10^{13}$ $^{\rm f}$ & \citet{Esplugues2013b} \\
\enddata
\tablenotetext{a}{Averaged column densities are estimated with the assumption of the source diameter of $15''$.}
\tablenotetext{b}{References for Orion~KL.}
\tablenotetext{c}{The column density is estimated by summing the column density of cold and hot component.}
\tablenotetext{d}{The column density is estimated under the LTE approximation with $T_{\rm ex} = 15$~K.}
\tablenotetext{e}{Following the method described by \citet{Fuente2014}, we estimate the averaged column density $N$ by summing up the column densities of extended ridge ($N_{\rm er}$), compact ridge ($N_{\rm cr}$), plateau ($N_{\rm p}$), and hot core ($N_{\rm hc}$).  A filling factor of 0.44 is assumed only for the hot core, since \citet{Tercero2010}, \citet{Esplugues2013a} and \citet{Marcelino2009} assumed the source diameter of the hot core to be $10''$ and the source size of other components to be larger than $15''$.  Namely, $N = N_{\rm er} + N_{\rm cr} + N_{\rm p} + 0.44 \times N_{\rm hc}$. }
\tablenotetext{f}{Following the method described by \citet{Fuente2014}, we estimate the averaged column density $N$ by summing up column densities of extended ridge ($N_{\rm er}$), compact ridge ($N_{\rm cr}$), plateau ($N_{\rm p}$), outer hot core ($N_{\rm ohc}$), and inner hot core ($N_{\rm ihc}$).  A filling factor of 0.44 and 0.22 is assumed for the outer hot core and the inner hot core, respectively, since \citet{Esplugues2013b} assumed the source diameter of outer hot core and inner hot core to be $10''$ and $7''$, respectively.  Namely, $N = N_{\rm er} + N_{\rm cr} + N_{\rm p} + 0.44 \times N_{\rm ohc} + 0.22 \times N_{\rm ihc}$. }
\tablenotetext{g}{Estimated from $^{13}$C bearing species with assumption that $^{12}$C/$^{13}$C = 50.}
\tablenotetext{h}{The column density estimated by \citet{Fuente2014}. }
\label{tab5}
\end{deluxetable}



\clearpage
\appendix
\setcounter{figure}{0}
 \renewcommand{\thefigure}{A.\arabic{figure}}
\section{Expanded Spectra}
\begin{figure}
\epsscale{1.00}
\plotone{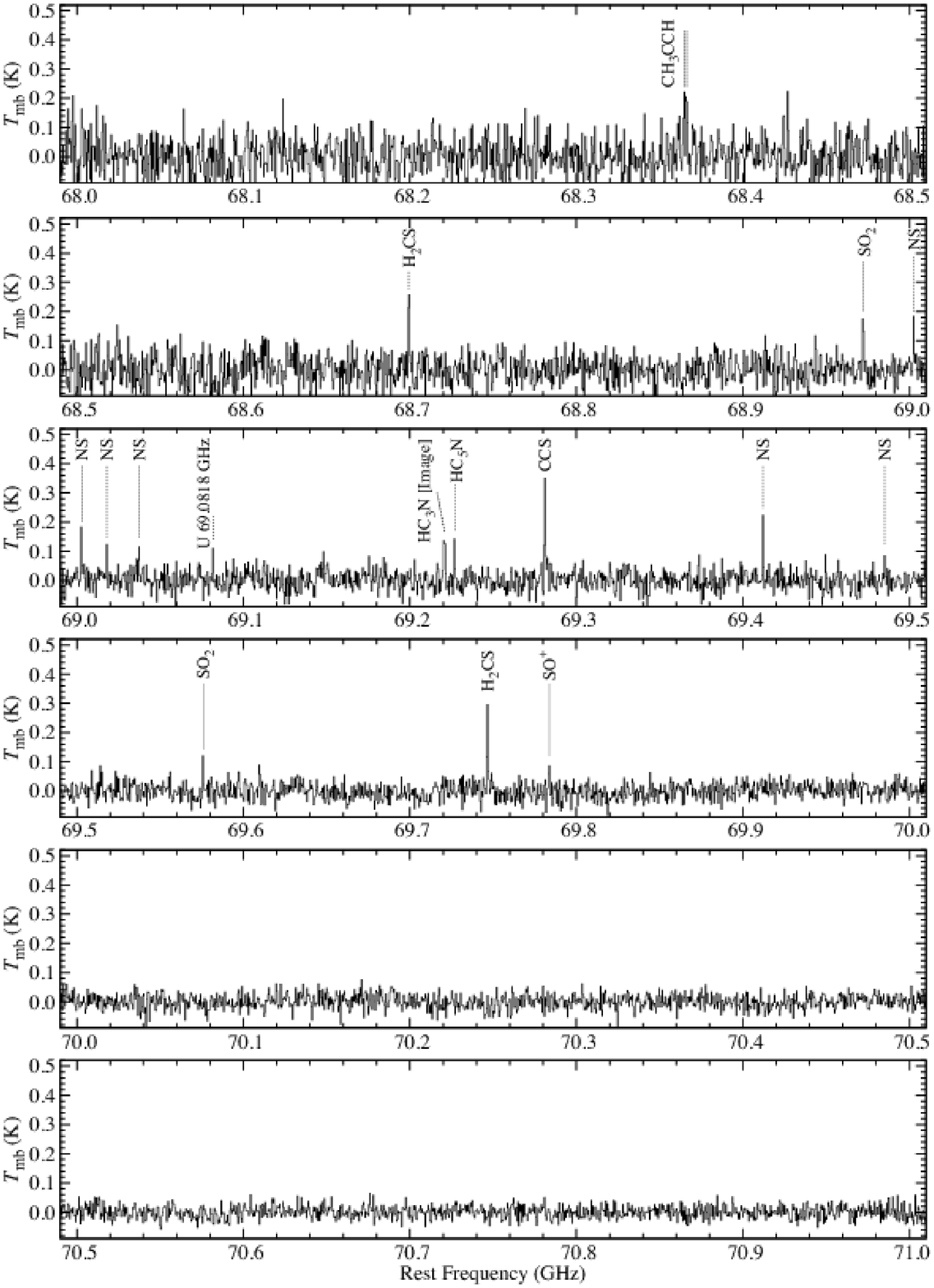}
\caption{Spectra of NGC~2264~CMM3 in the 3 and 4~mm band.  The frequency resolution is 0.5~MHz.  The $V_{\rm LSR}$ is assumed assumed to be 7.0~km~s$^{-1}$.}
\label{figa1}
\end{figure}
\setcounter{figure}{0}

\begin{figure}
\epsscale{1.00}
\plotone{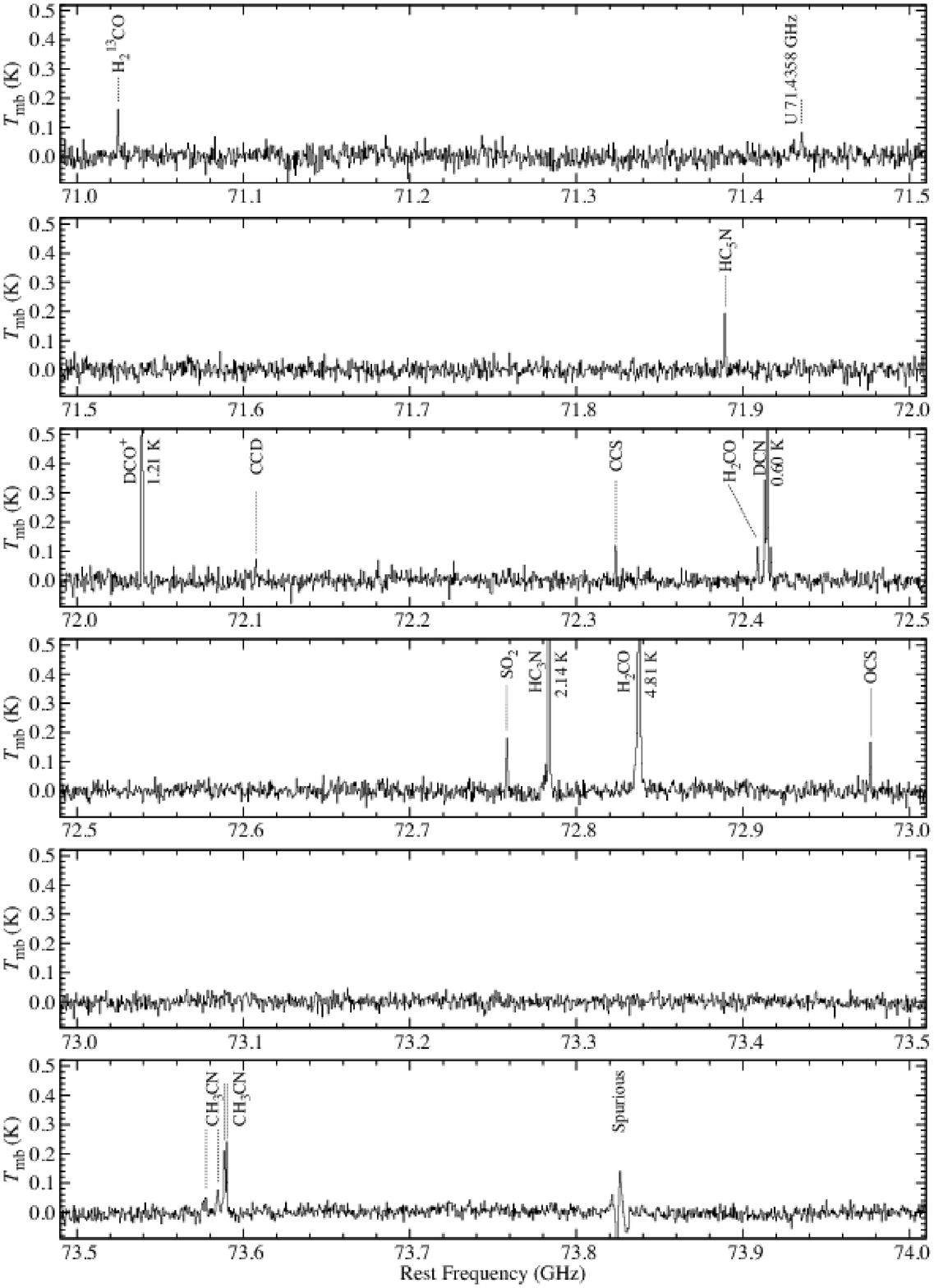}
\caption{\textit{Continued}}
\end{figure}
\setcounter{figure}{0}

\begin{figure}
\epsscale{1.00}
\plotone{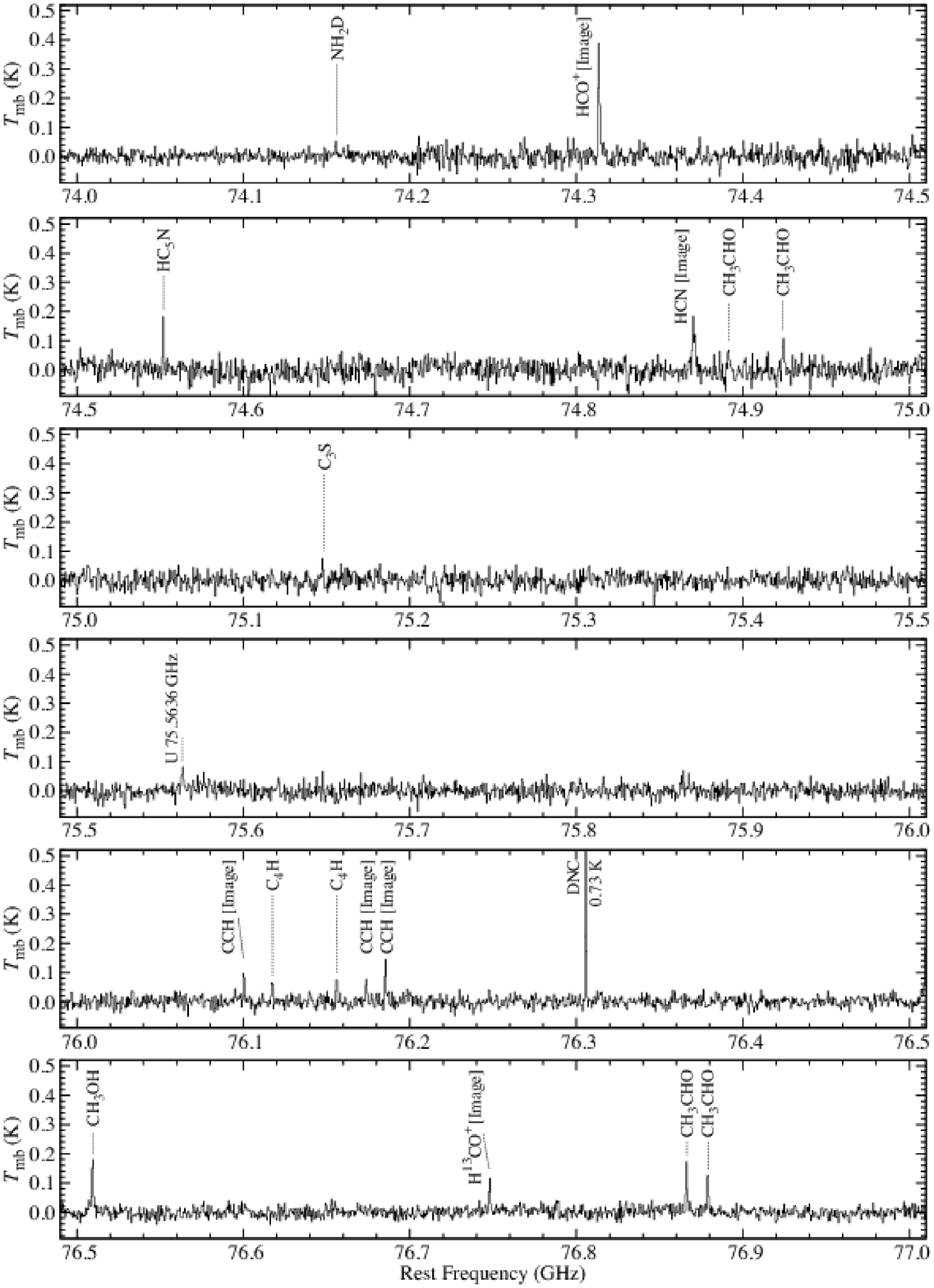}
\caption{\textit{Continued}}
\end{figure}
\setcounter{figure}{0}

\begin{figure}
\epsscale{1.00}
\plotone{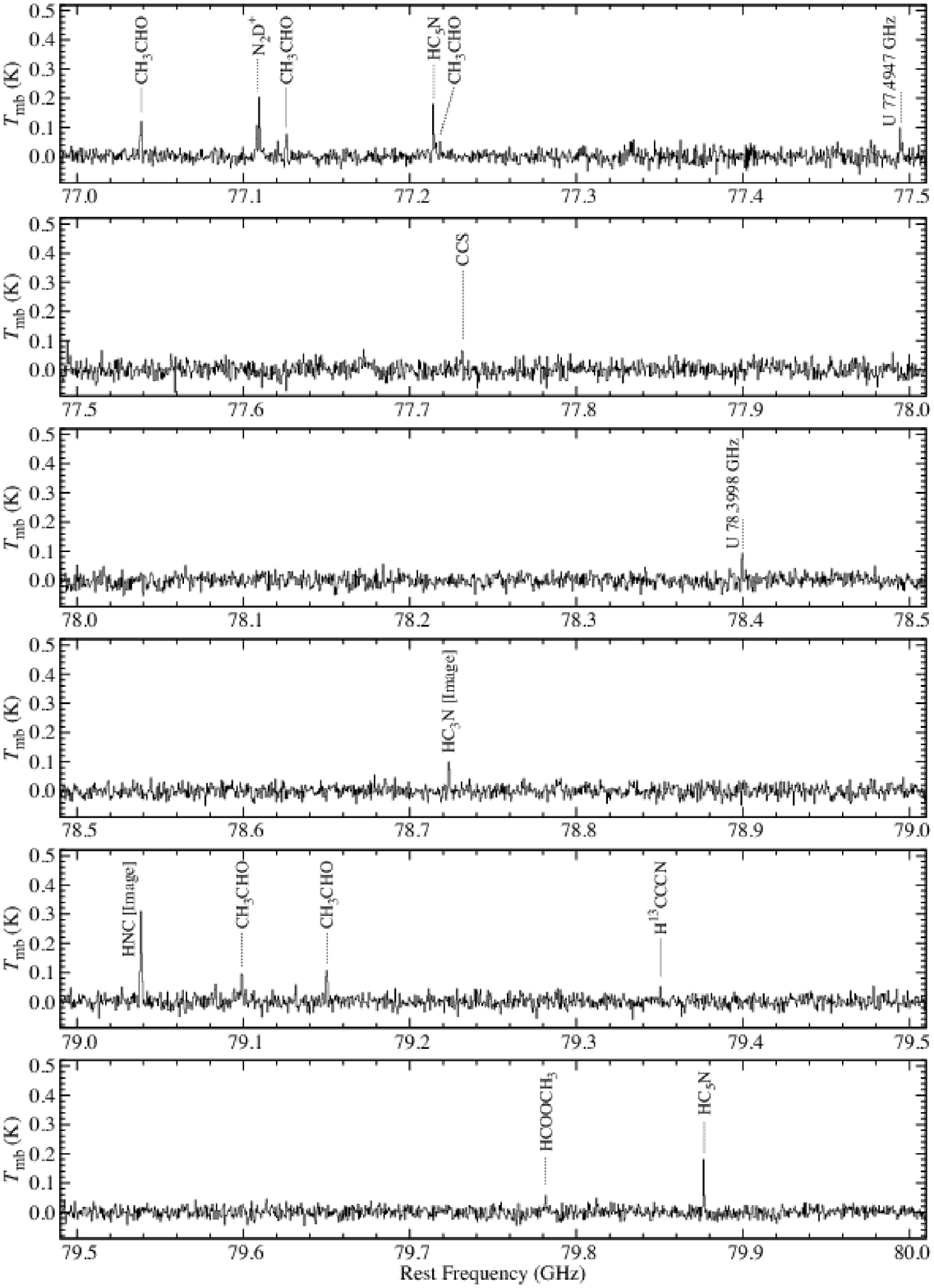}
\caption{\textit{Continued}}
\end{figure}
\setcounter{figure}{0}

\begin{figure}
\epsscale{1.00}
\plotone{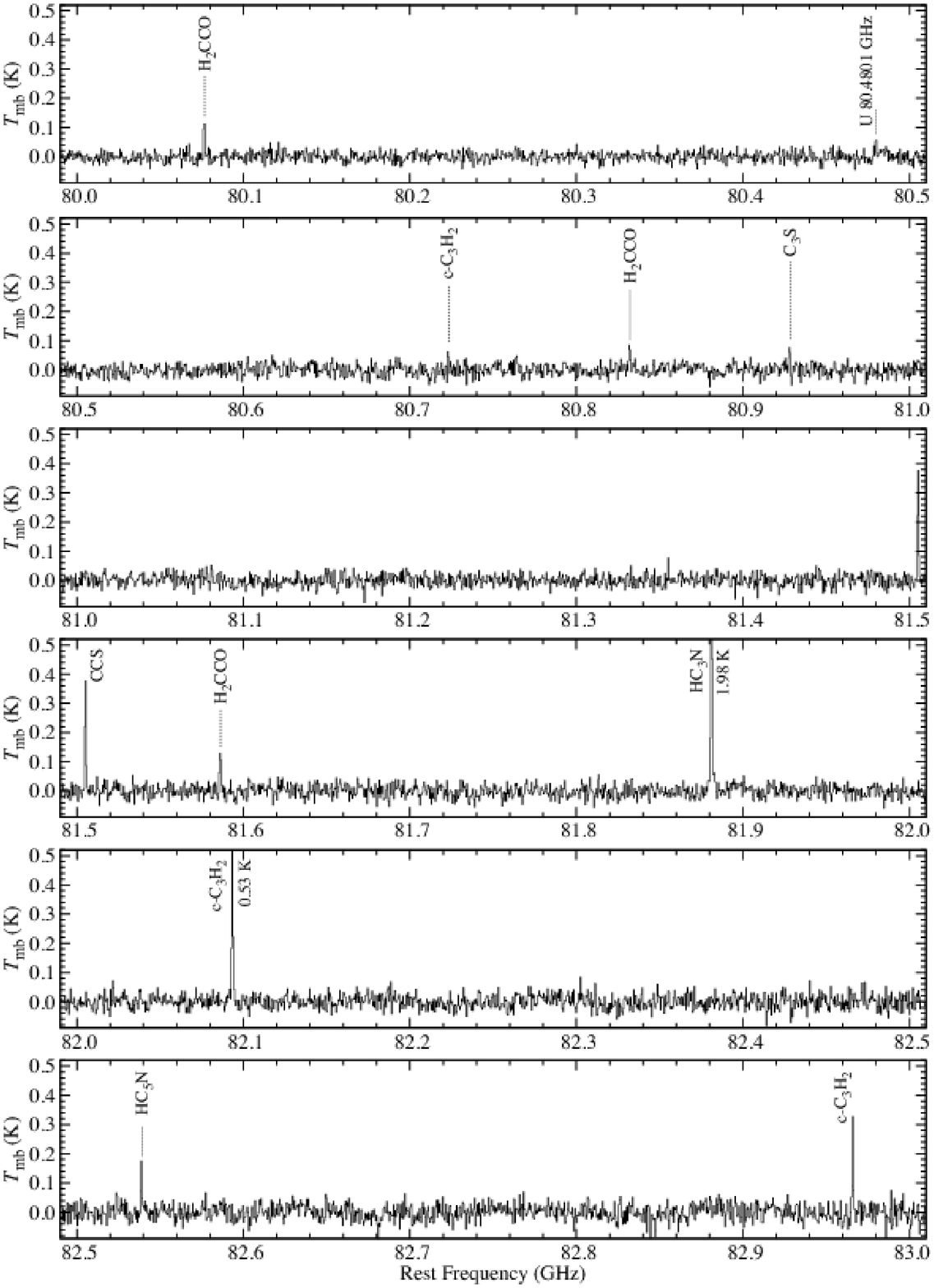}
\caption{\textit{Continued}}
\setcounter{figure}{0}
\end{figure}

\begin{figure}
\epsscale{1.00}
\plotone{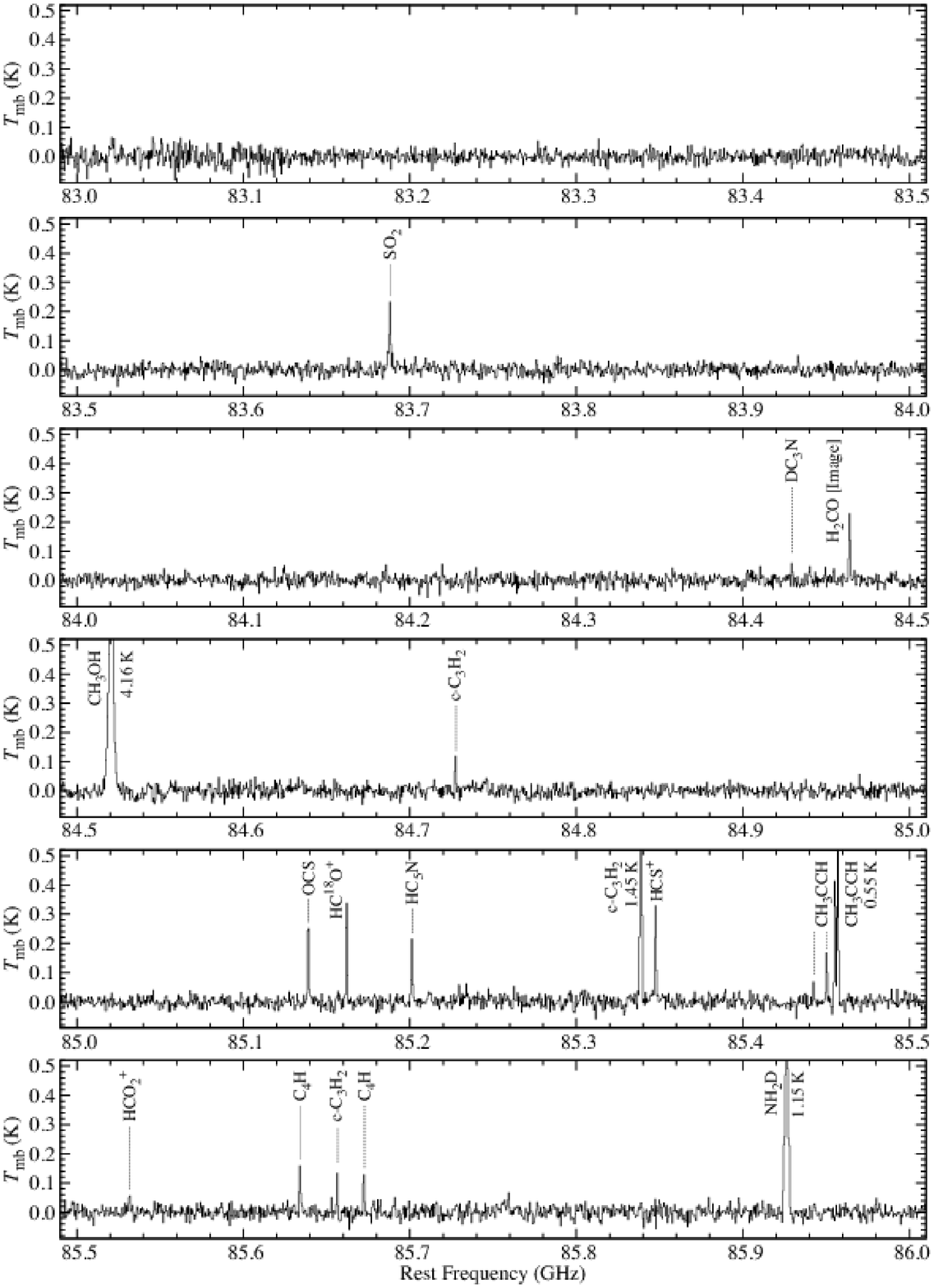}
\caption{\textit{Continued}}
\end{figure}
\setcounter{figure}{0}

\begin{figure}
\epsscale{1.00}
\plotone{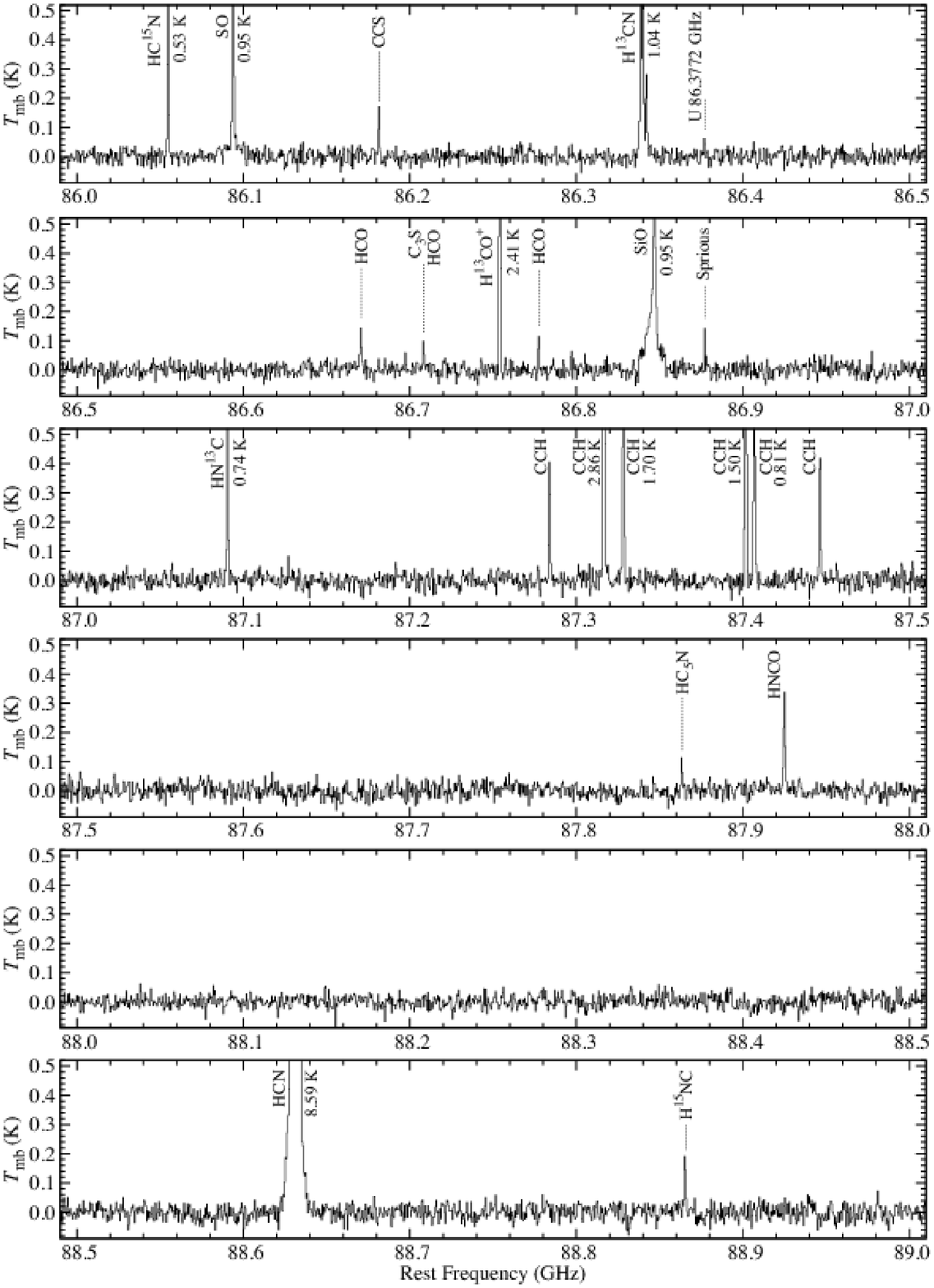}
\caption{\textit{Continued}}
\end{figure}
\setcounter{figure}{0}

\begin{figure}
\epsscale{1.00}
\plotone{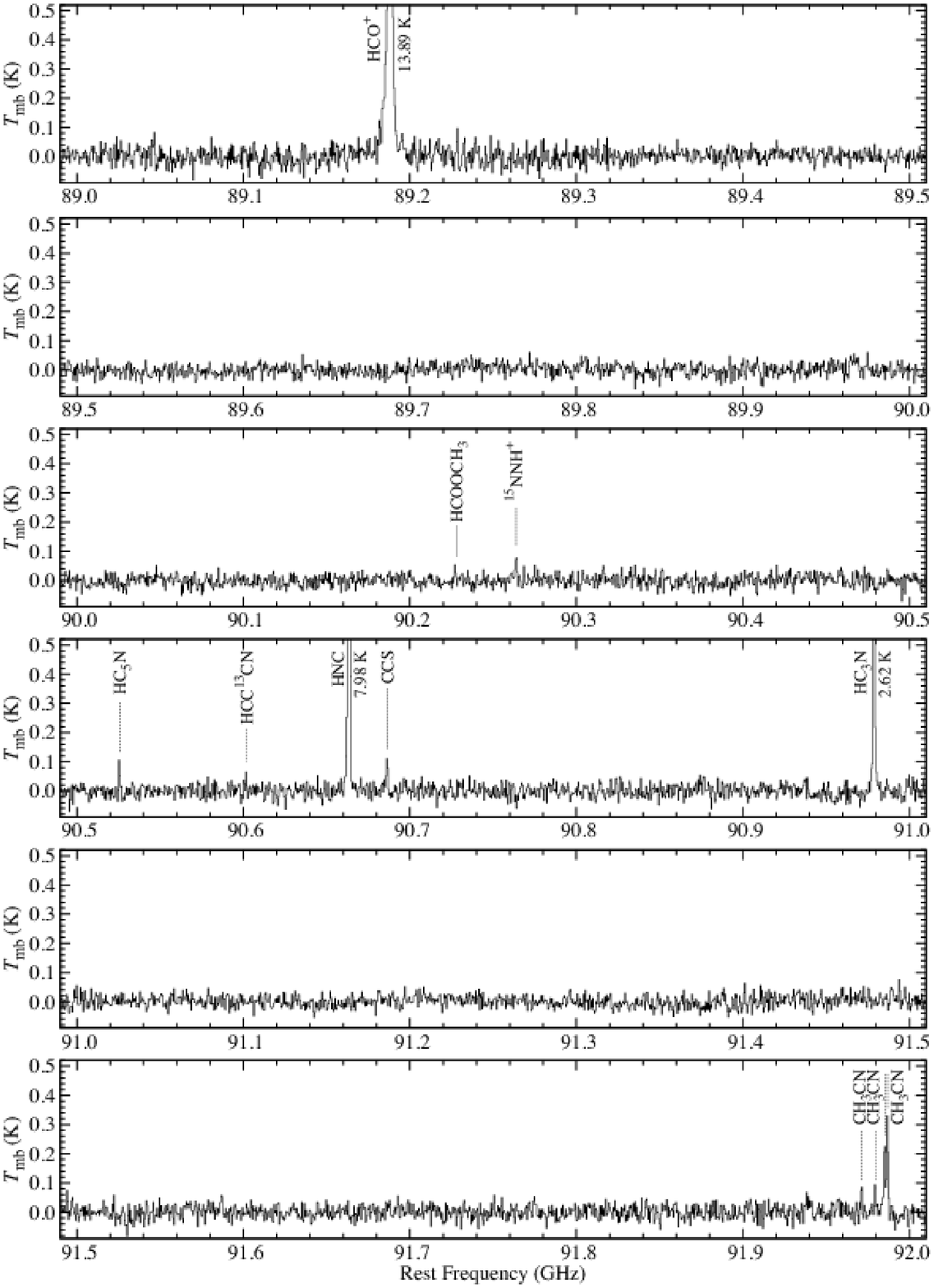}
\caption{\textit{Continued}}
\end{figure}
\setcounter{figure}{0}

\begin{figure}
\epsscale{1.00}
\plotone{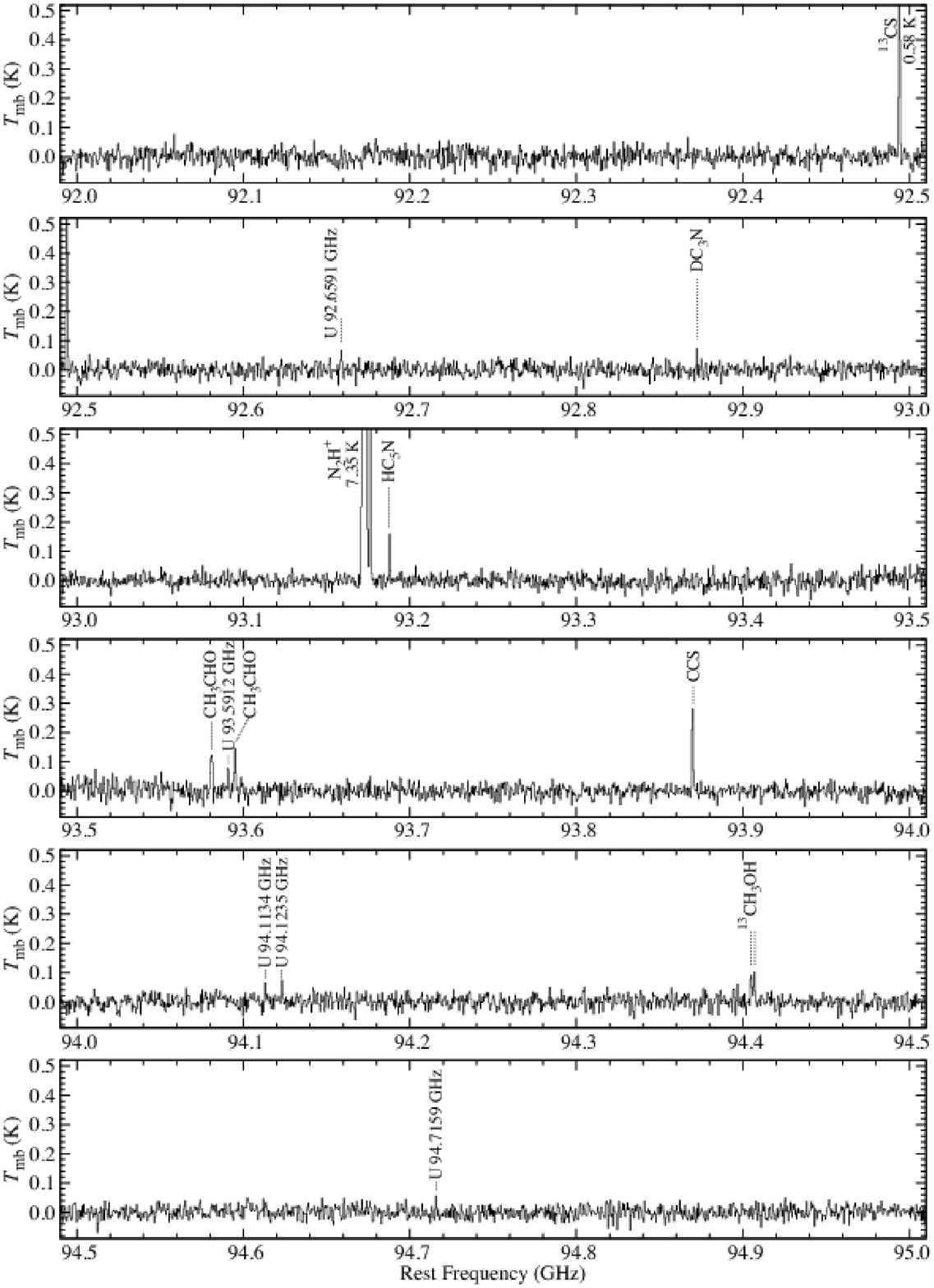}
\caption{\textit{Continued}}
\end{figure}
\setcounter{figure}{0}

\begin{figure}
\epsscale{1.00}
\plotone{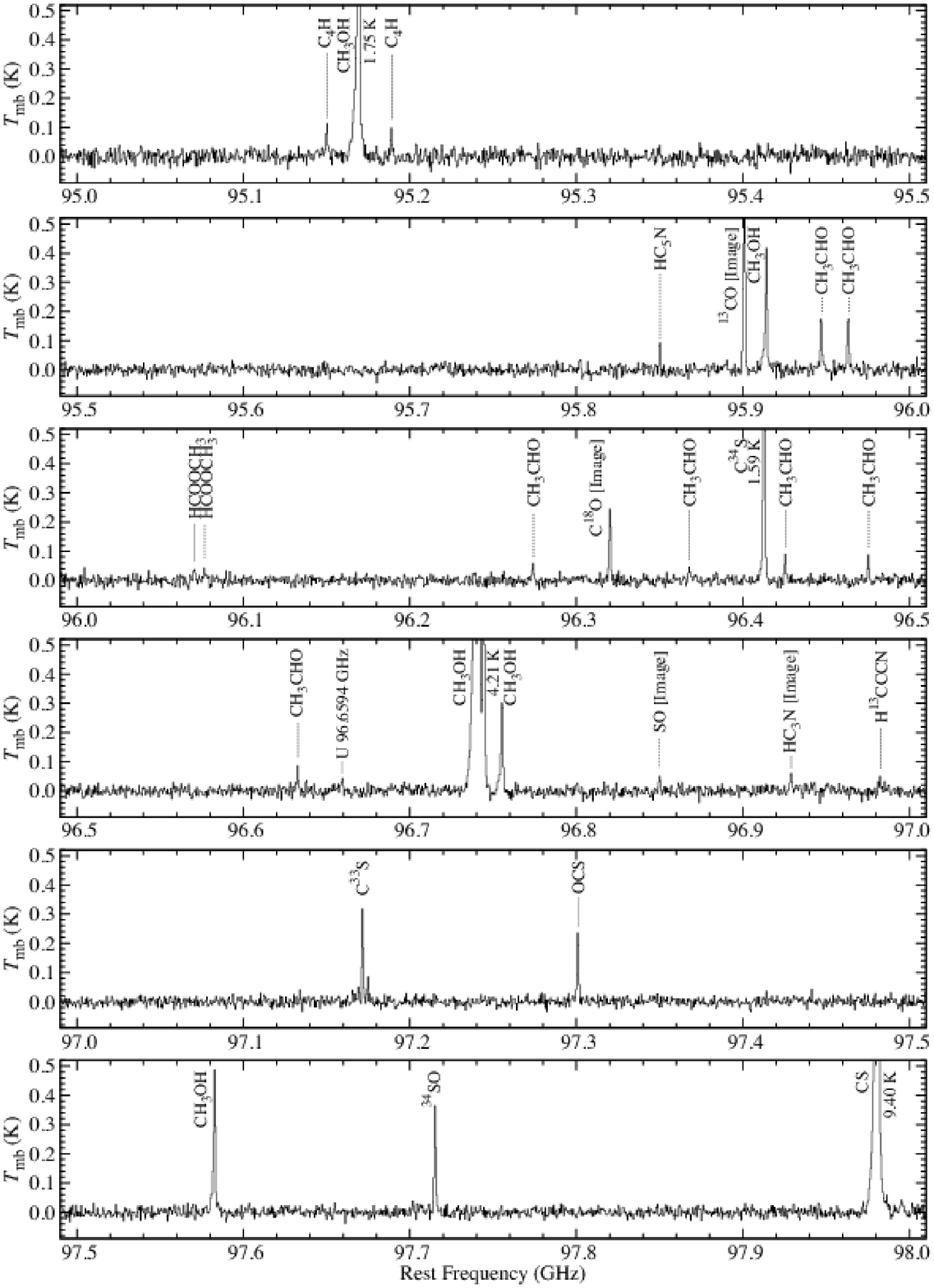}
\caption{\textit{Continued}}
\end{figure}
\setcounter{figure}{0}

\begin{figure}
\epsscale{1.00}
\plotone{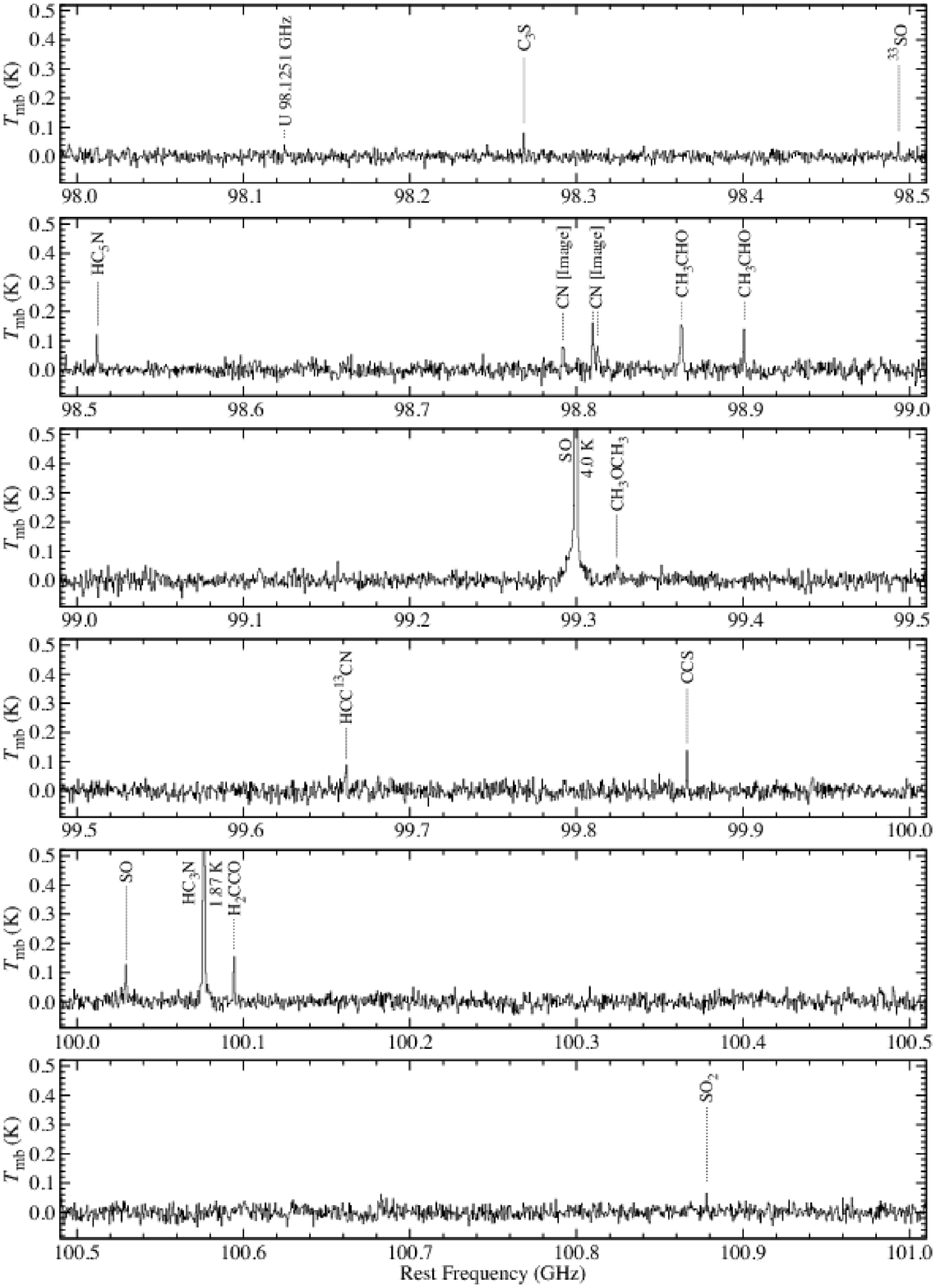}
\caption{\textit{Continued}}
\end{figure}
\setcounter{figure}{0}

\begin{figure}
\epsscale{1.00}
\plotone{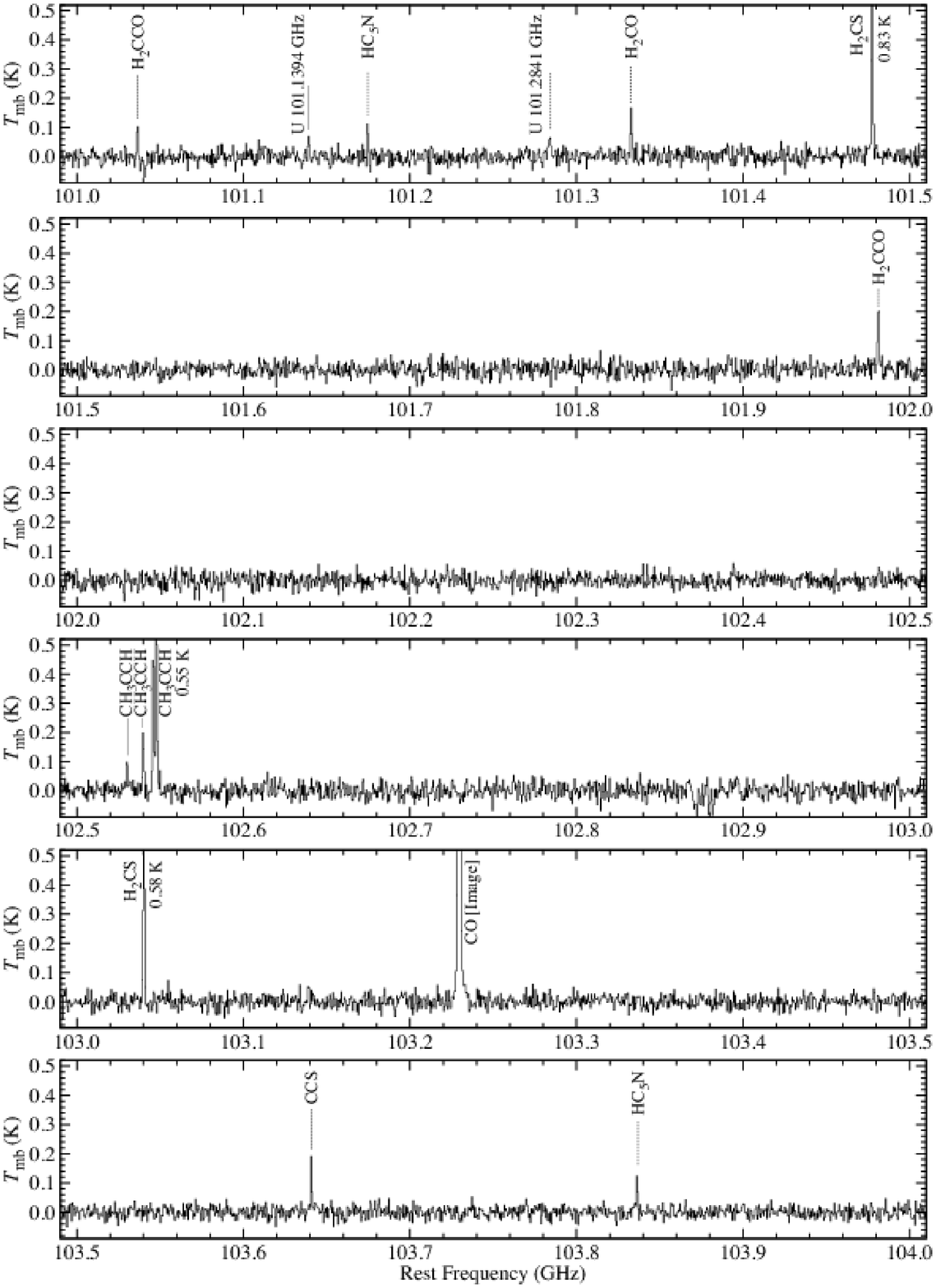}
\caption{\textit{Continued}}
\end{figure}
\setcounter{figure}{0}

\begin{figure}
\epsscale{1.00}
\plotone{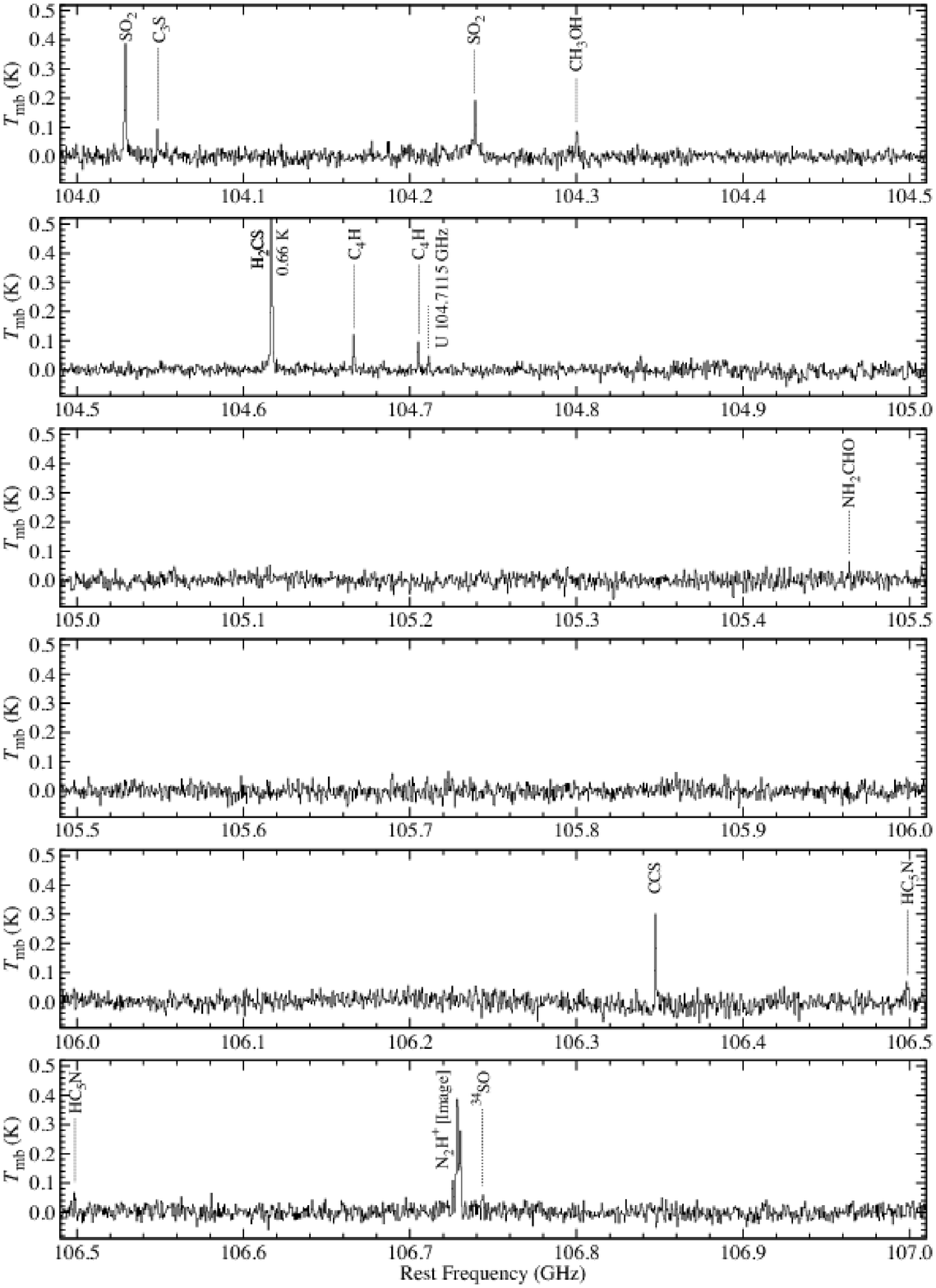}
\caption{\textit{Continued}}
\end{figure}
\setcounter{figure}{0}

\begin{figure}
\epsscale{1.00}
\plotone{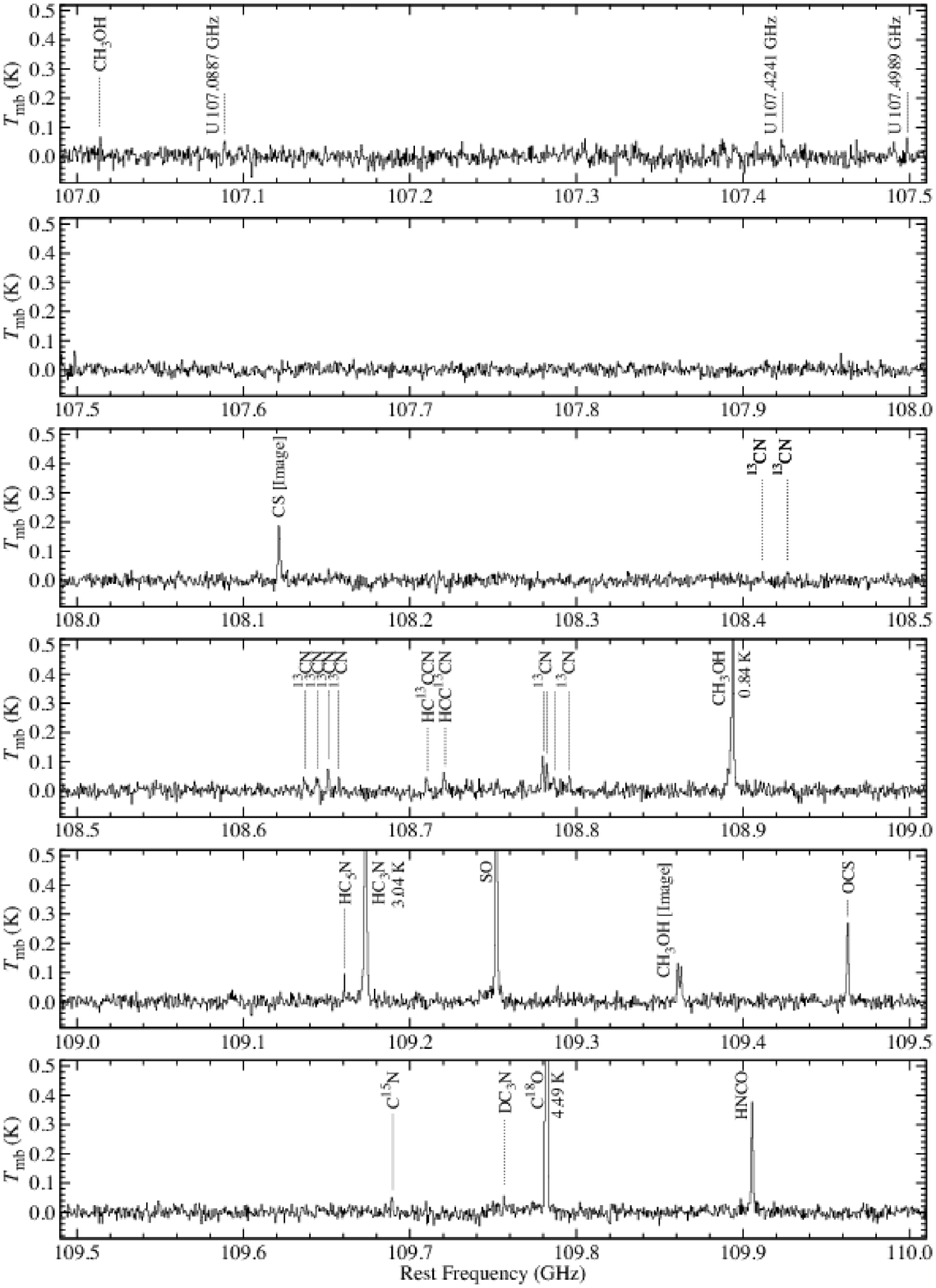}
\caption{\textit{Continued}}
\end{figure}
\setcounter{figure}{0}

\begin{figure}
\epsscale{1.00}
\plotone{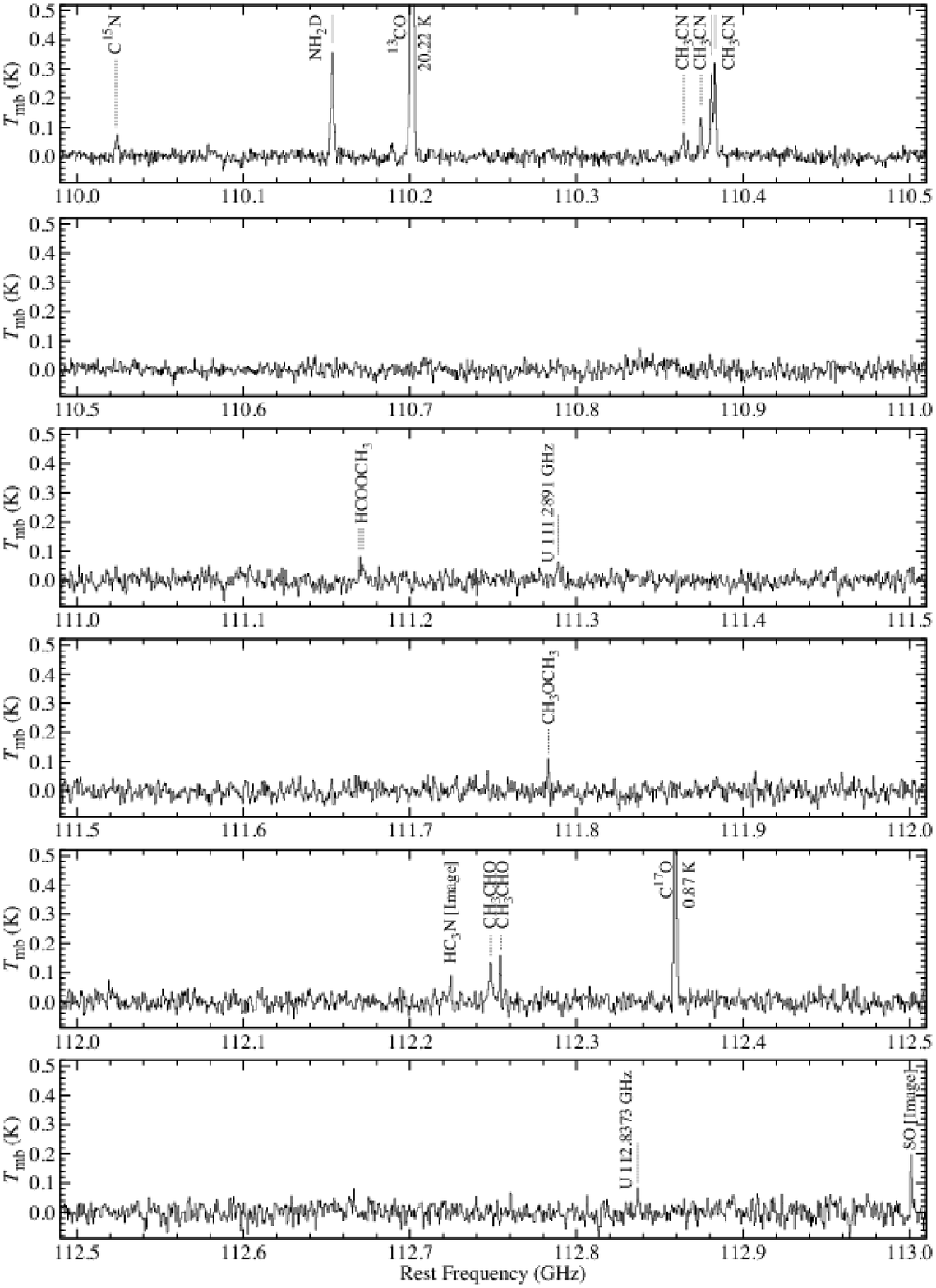}
\caption{\textit{Continued}}
\end{figure}
\setcounter{figure}{0}

\begin{figure}
\epsscale{1.00}
\plotone{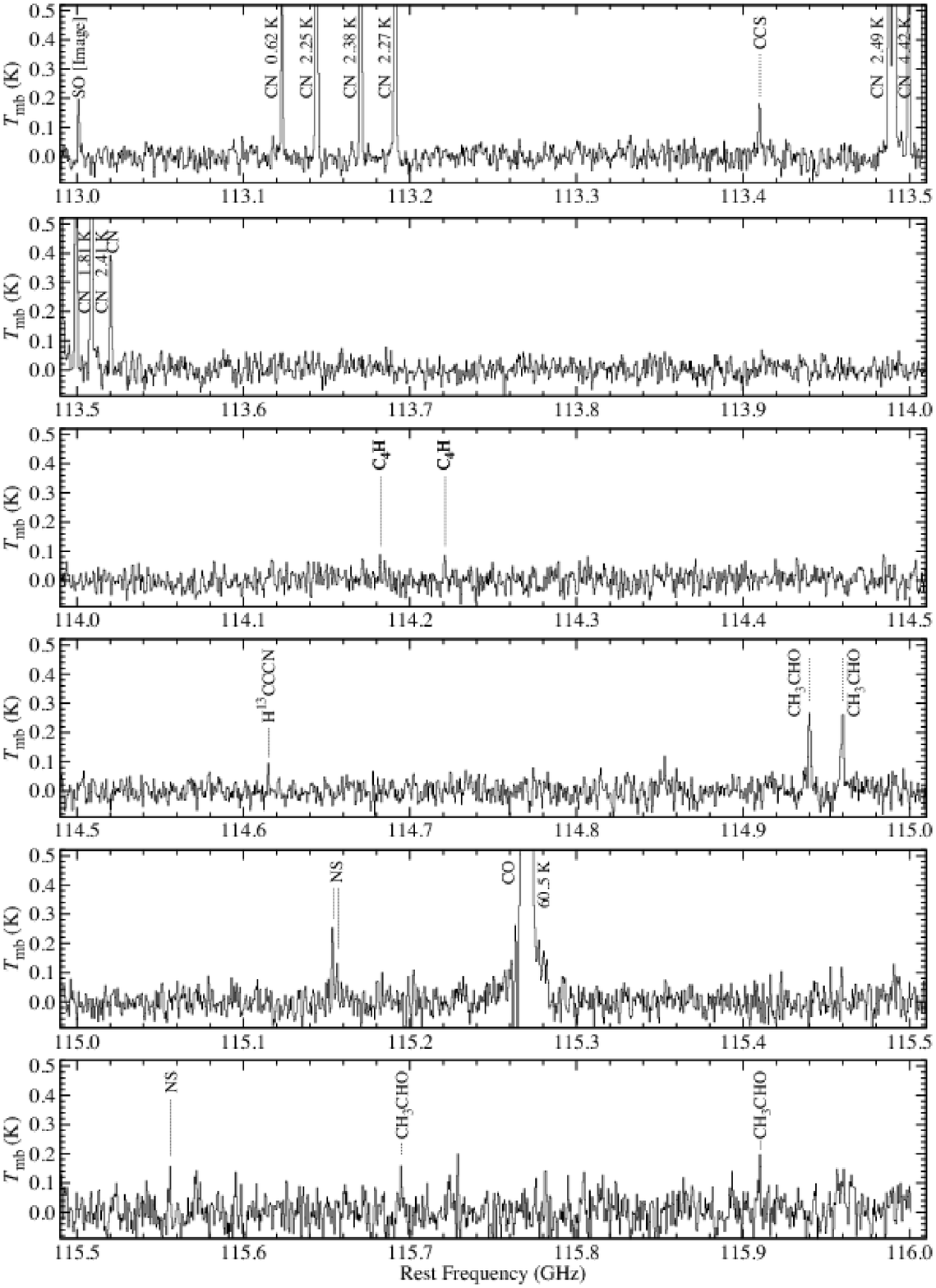}
\caption{\textit{Continued}}
\end{figure}

\clearpage 

\begin{figure}
\epsscale{1.00}
\plotone{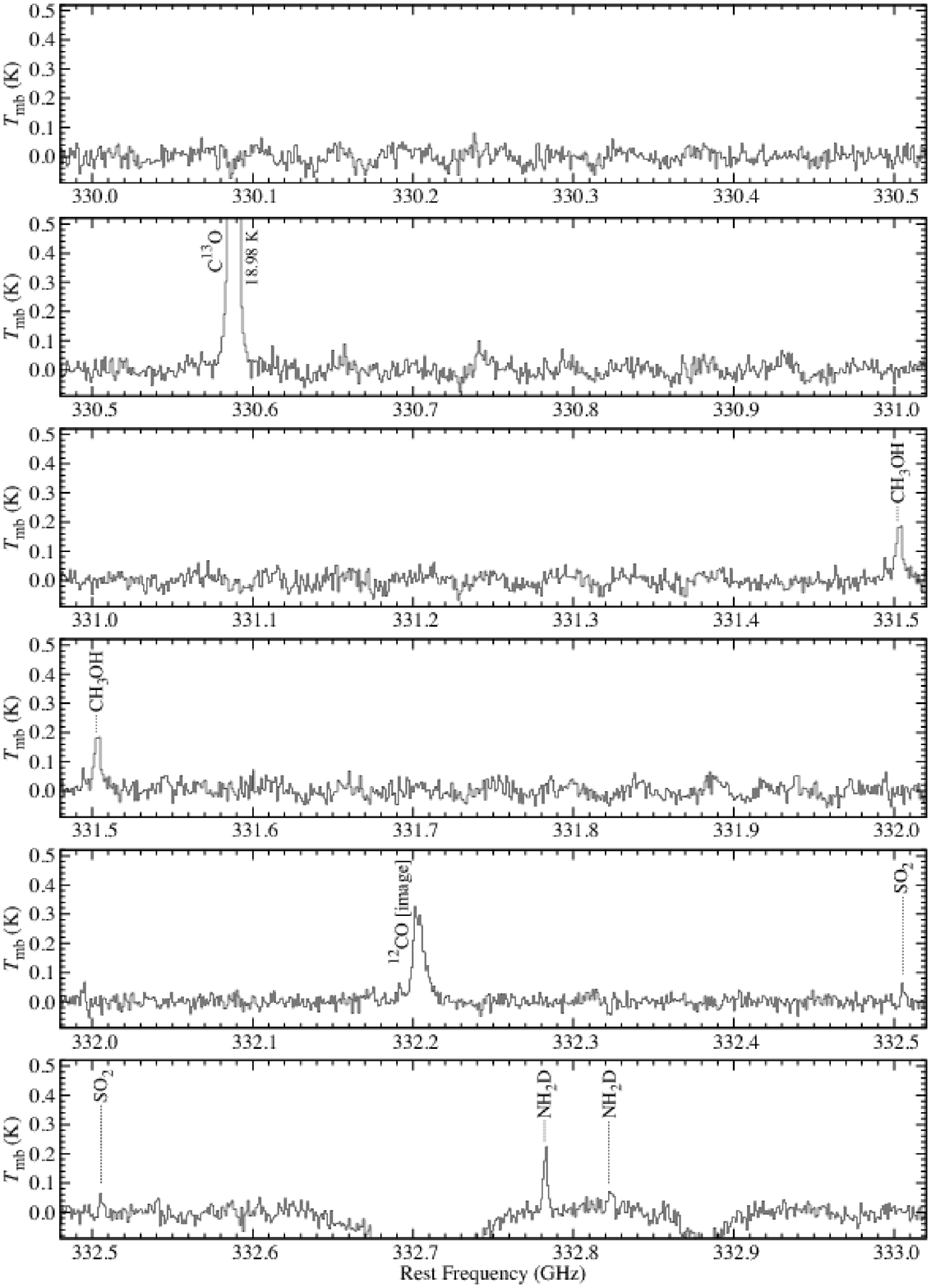}
\caption{Spectra of NGC~2264~CMM3 in the 0.8~mm band.  The frequency resolution is 1~MHz.  The $V_{\rm LSR}$ is assumed assumed to be 7.0~km~s$^{-1}$.  (O$_3$) indicates telluric ozone.}
\label{figa2}
\end{figure}
\setcounter{figure}{1}

\begin{figure}
\epsscale{1.00}
\plotone{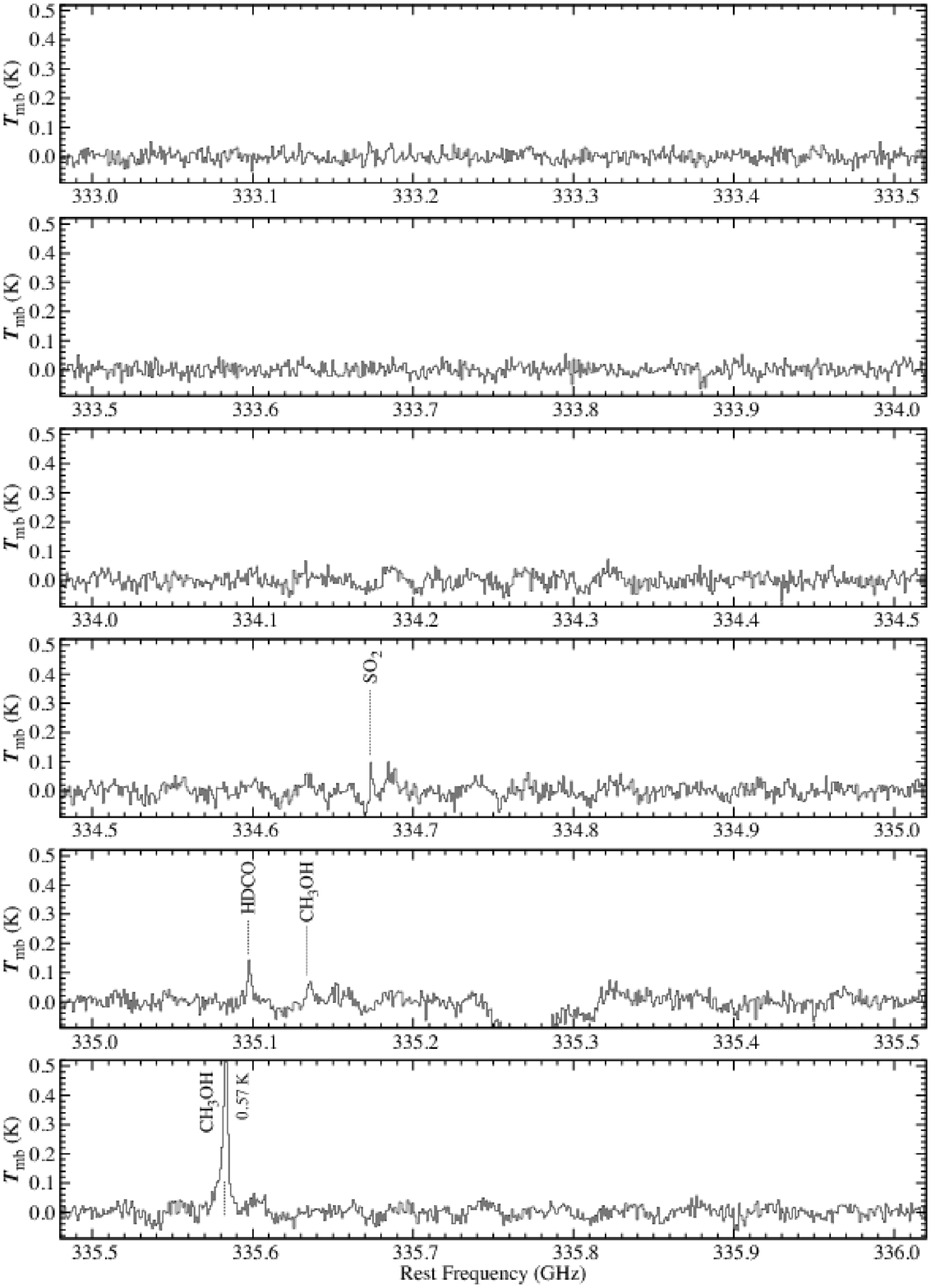}
\caption{\textit{Continued}}
\end{figure}
\setcounter{figure}{1}

\begin{figure}
\epsscale{1.00}
\plotone{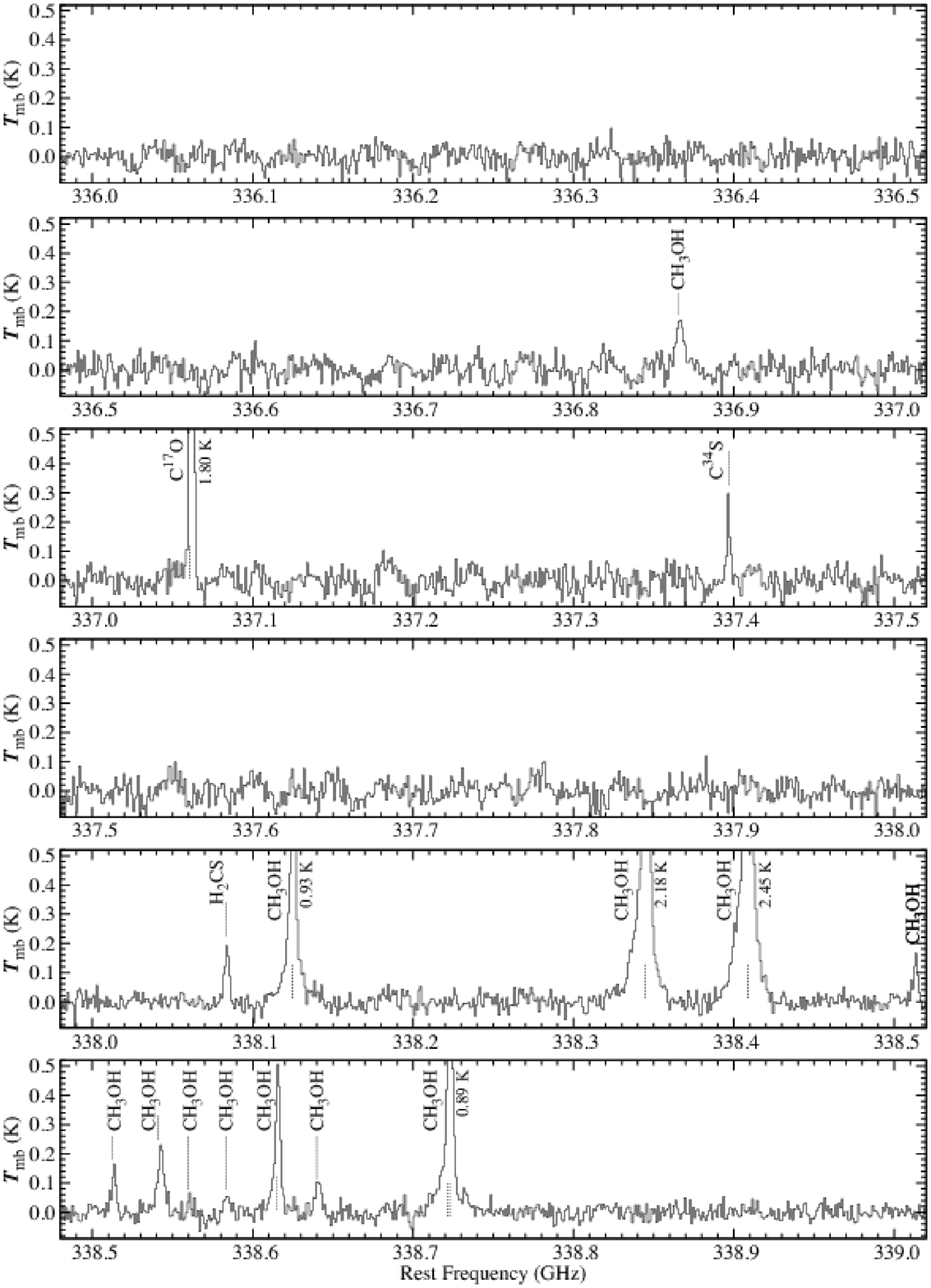}
\caption{\textit{Continued}}
\end{figure}
\setcounter{figure}{1}

\begin{figure}
\epsscale{1.00}
\plotone{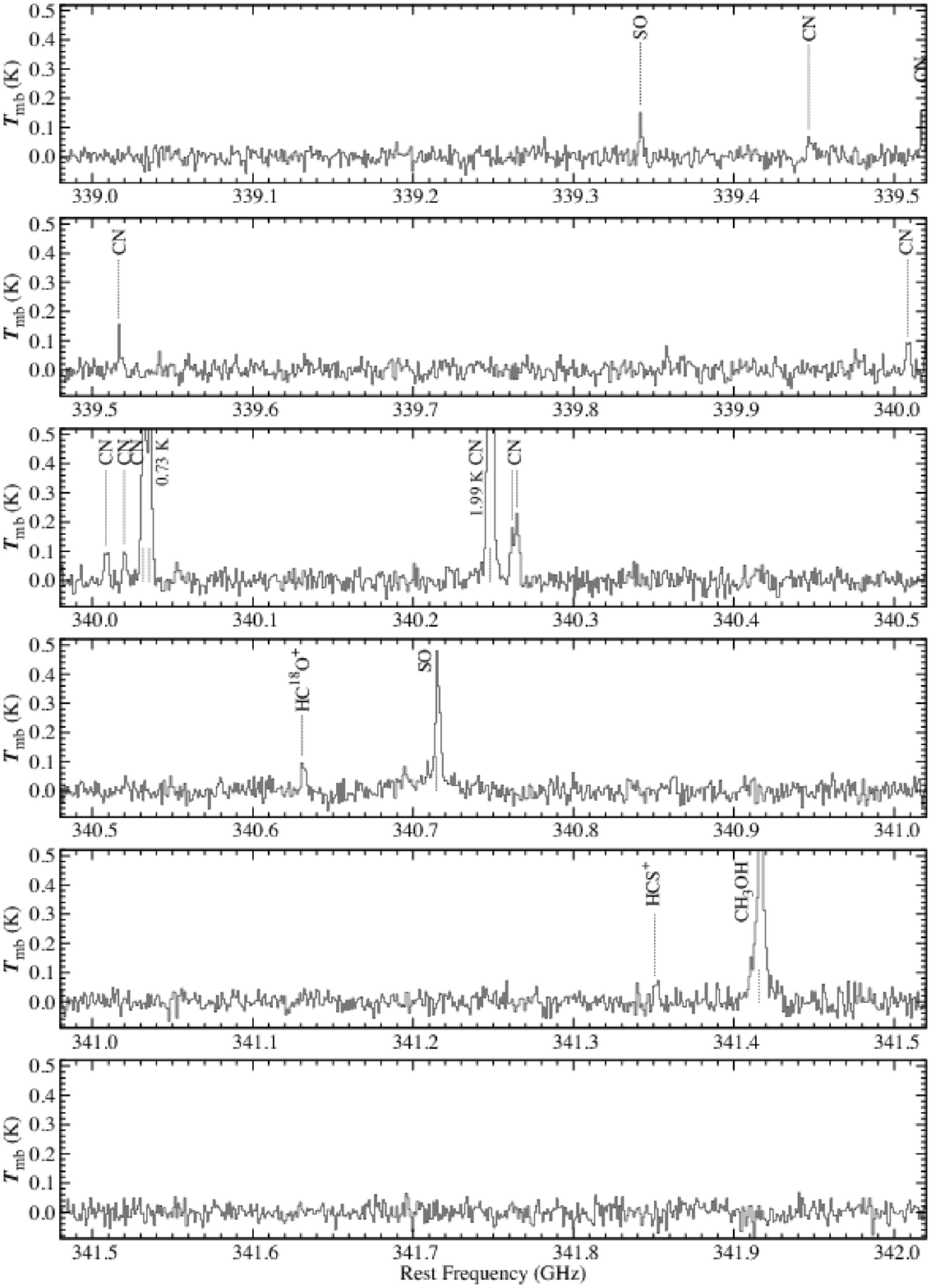}
\caption{\textit{Continued}}
\end{figure}
\setcounter{figure}{1}

\begin{figure}
\epsscale{1.00}
\plotone{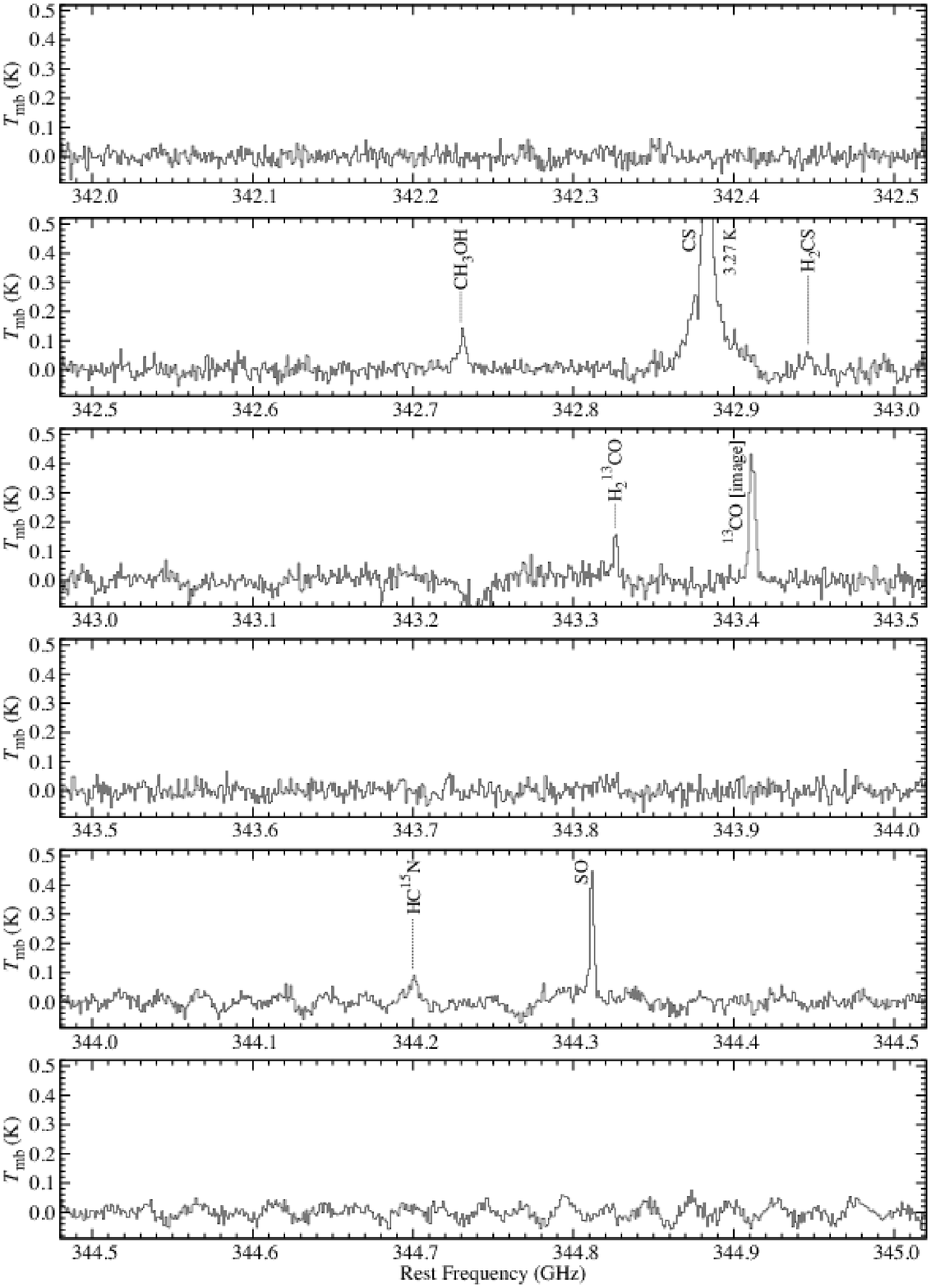}
\caption{\textit{Continued}}
\end{figure}
\setcounter{figure}{1}

\begin{figure}
\epsscale{1.00}
\plotone{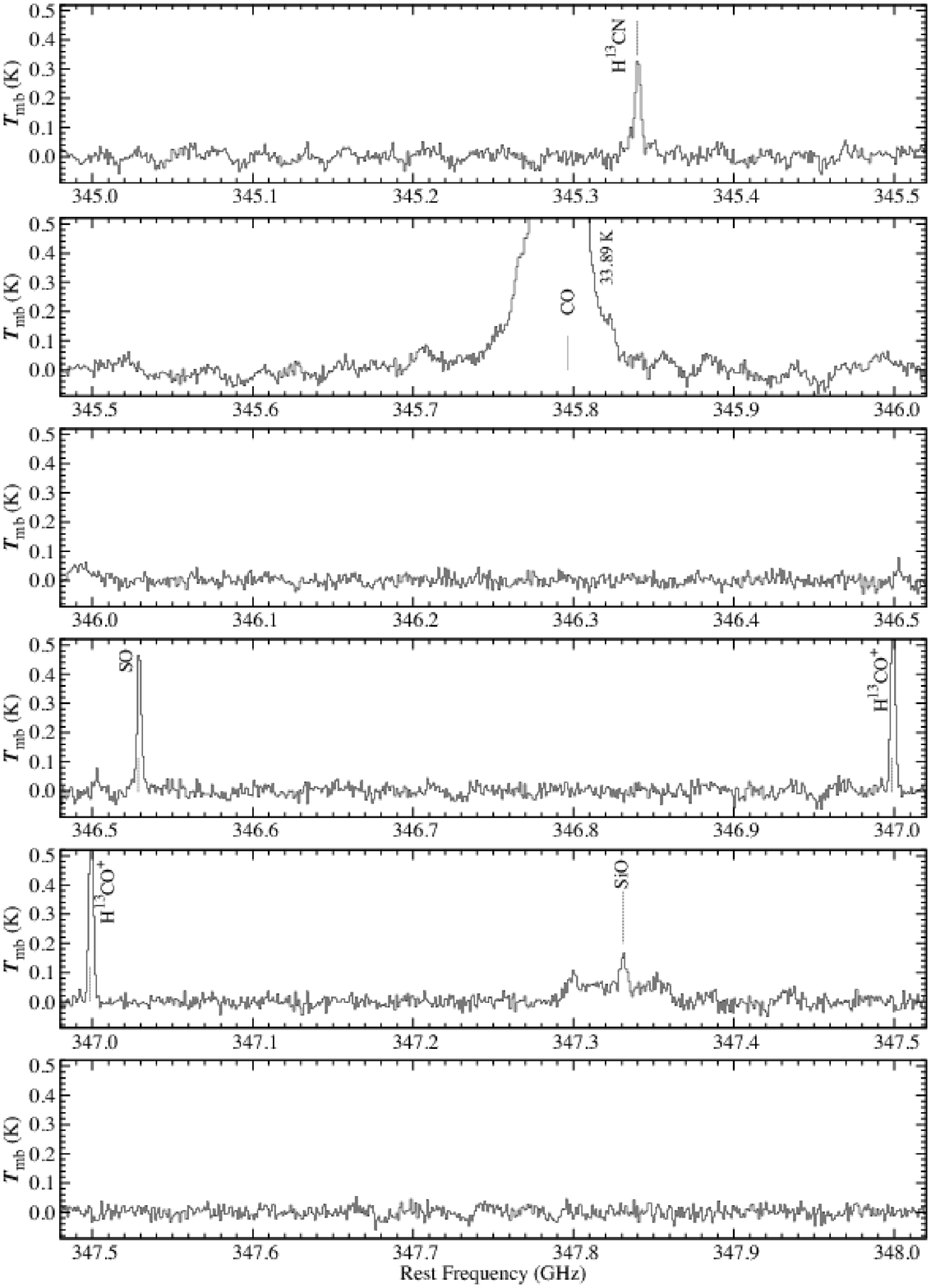}
\caption{\textit{Continued}}
\end{figure}
\setcounter{figure}{1}

\begin{figure}
\epsscale{1.00}
\plotone{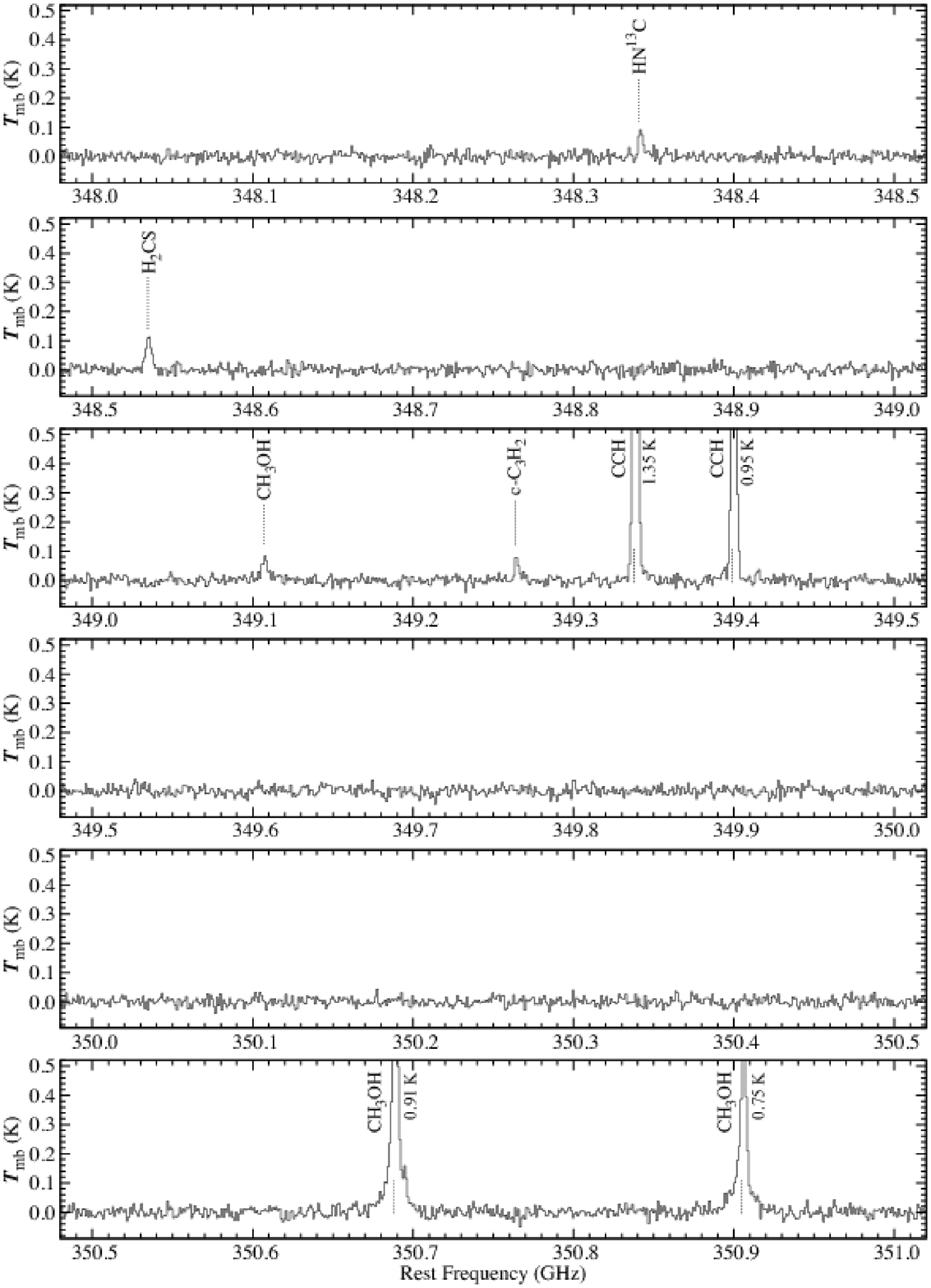}
\caption{\textit{Continued}}
\end{figure}
\setcounter{figure}{1}

\begin{figure}
\epsscale{1.00}
\plotone{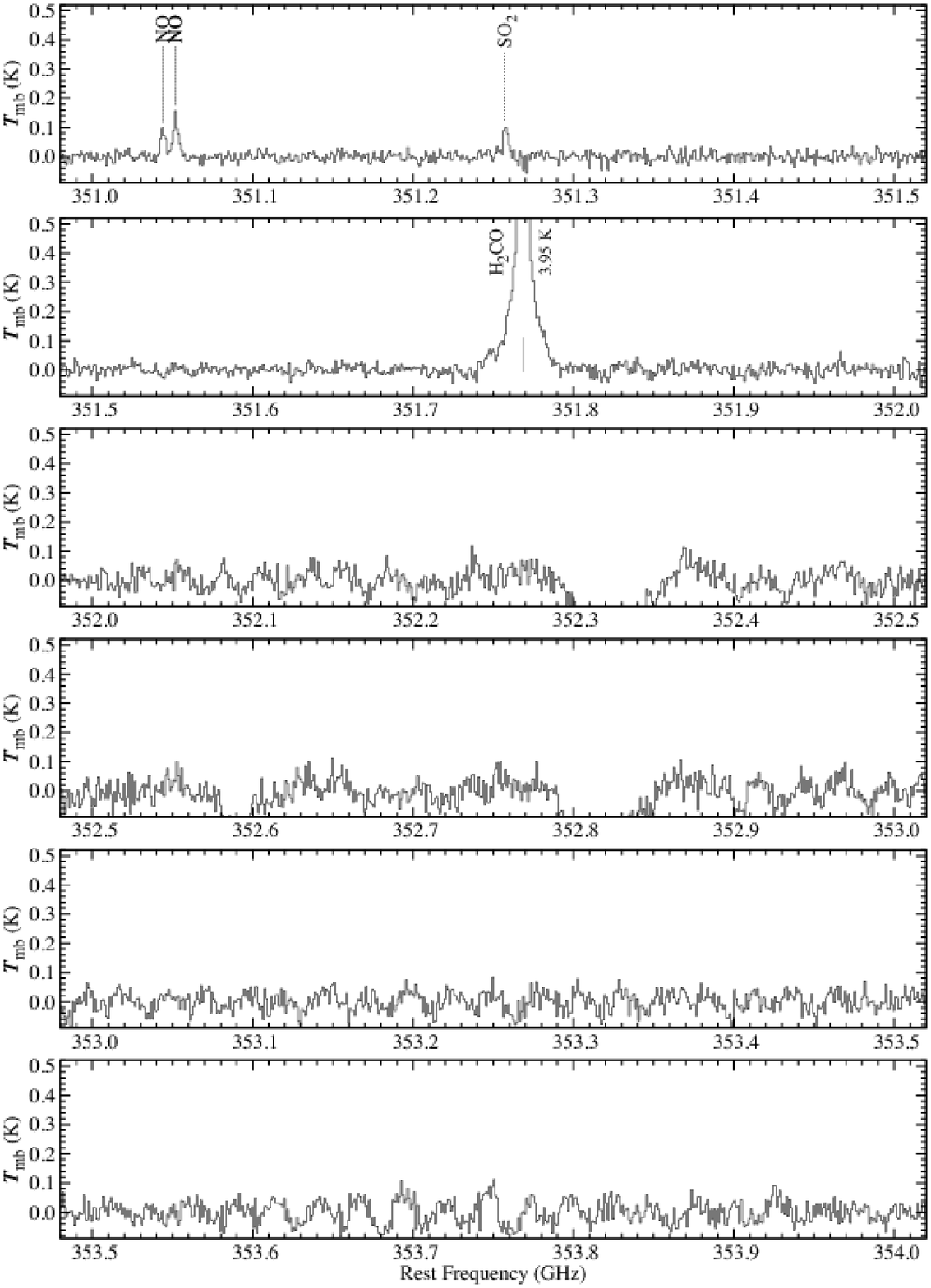}
\caption{\textit{Continued}}
\end{figure}
\setcounter{figure}{1}

\begin{figure}
\epsscale{1.00}
\plotone{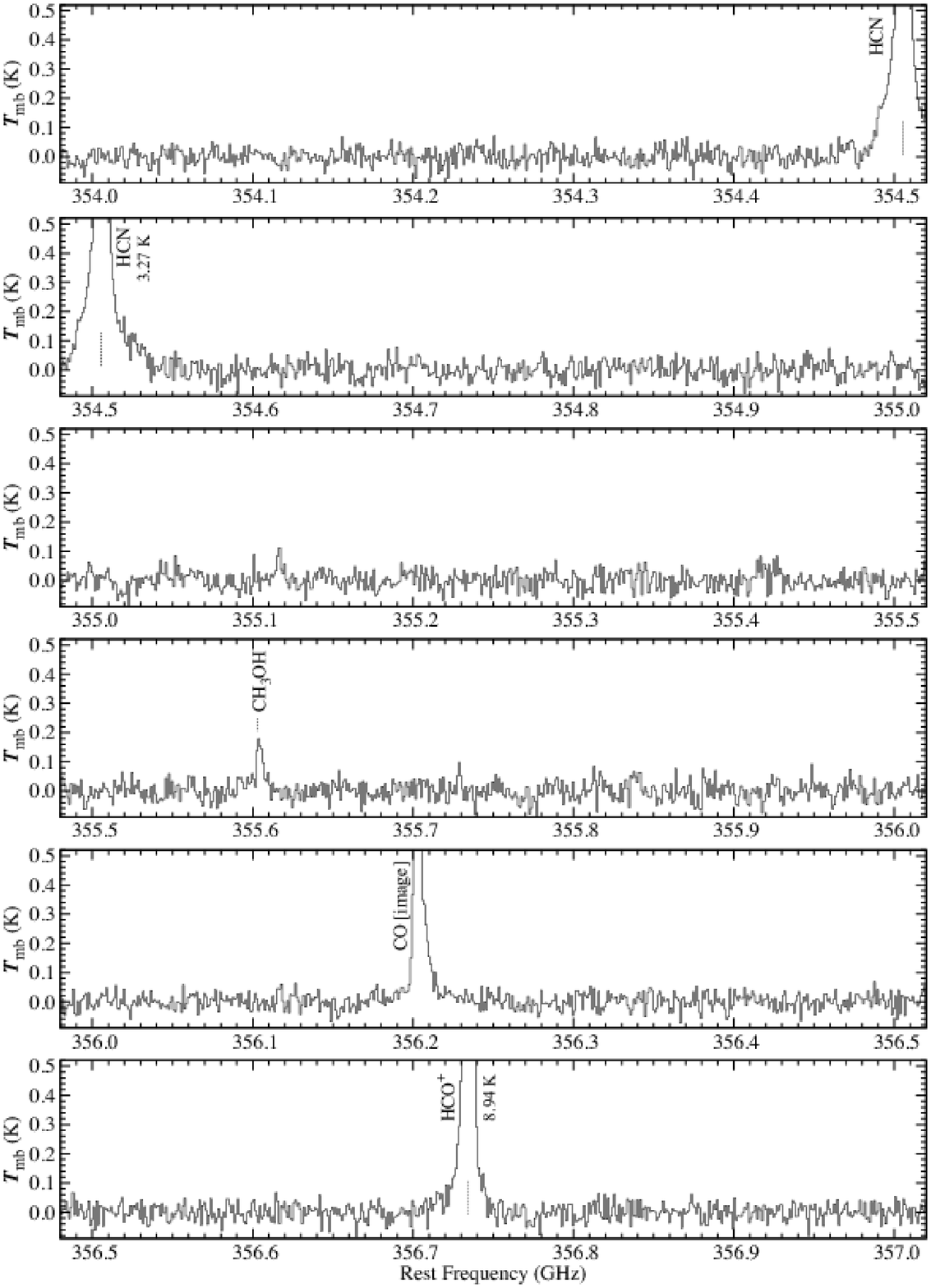}
\caption{\textit{Continued}}
\end{figure}
\setcounter{figure}{1}

\begin{figure}
\epsscale{1.00}
\plotone{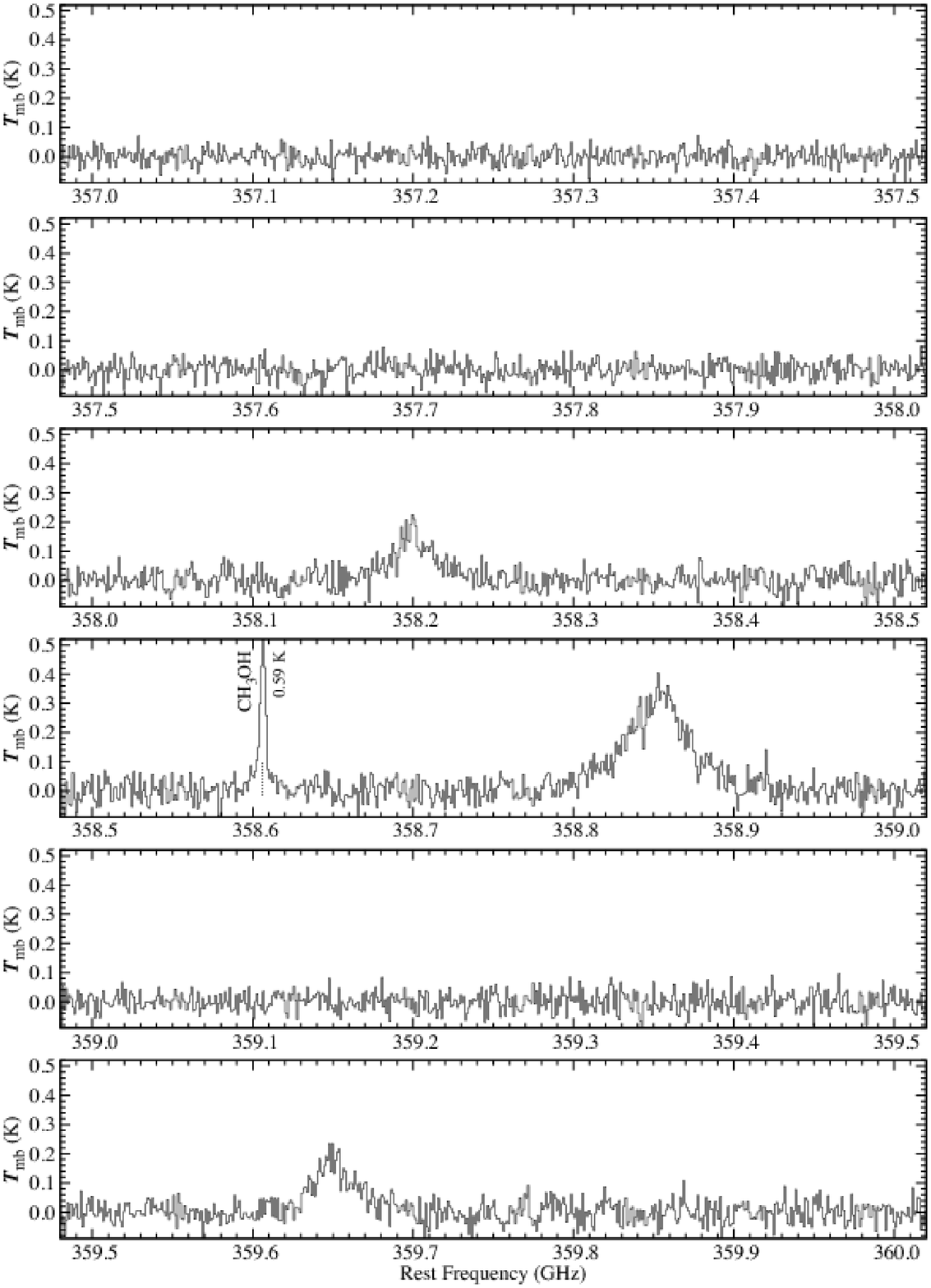}
\caption{\textit{Continued}}
\end{figure}
\setcounter{figure}{1}

\begin{figure}
\epsscale{1.00}
\plotone{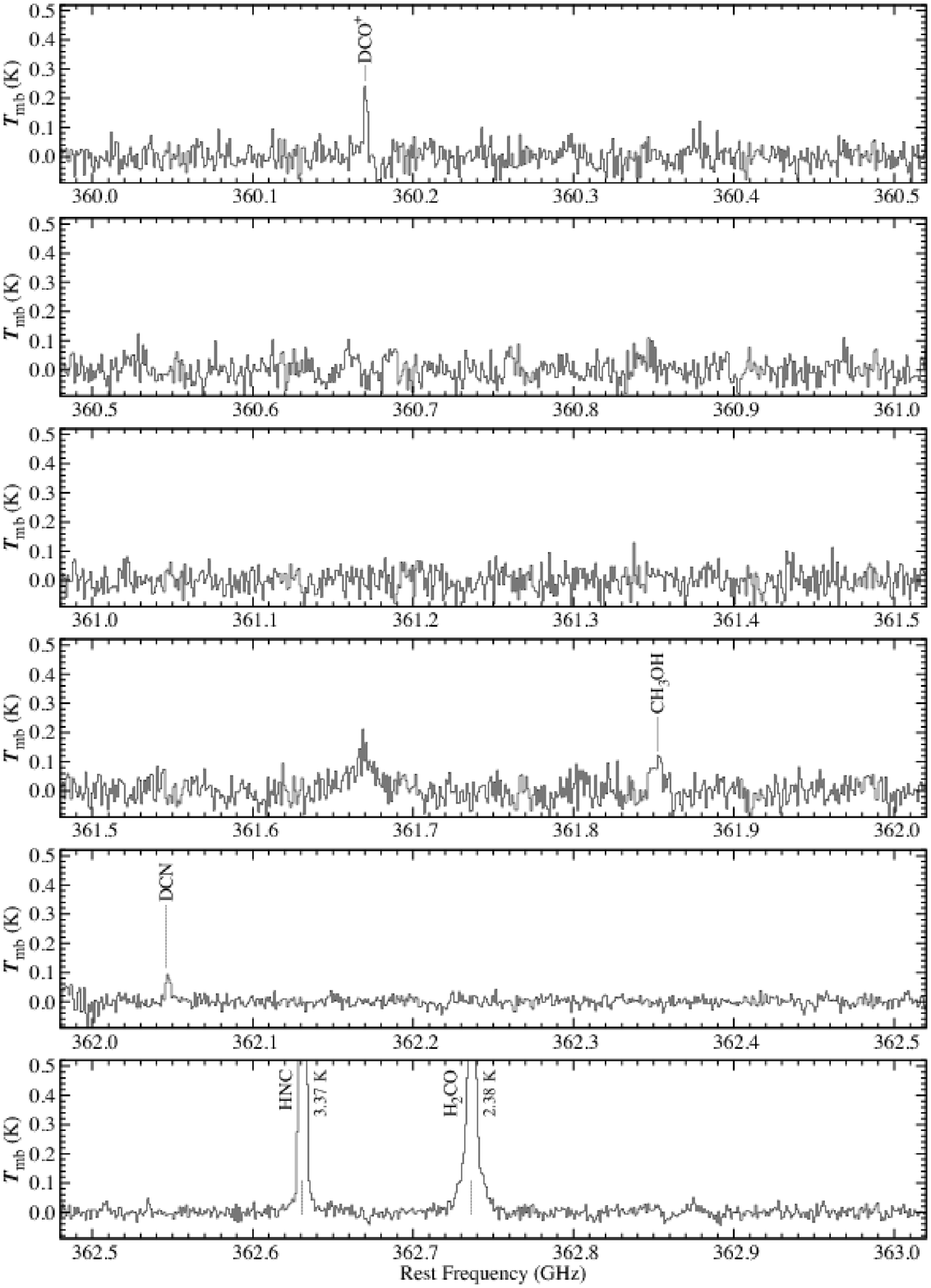}
\caption{\textit{Continued}}
\end{figure}
\setcounter{figure}{1}

\begin{figure}
\epsscale{1.00}
\plotone{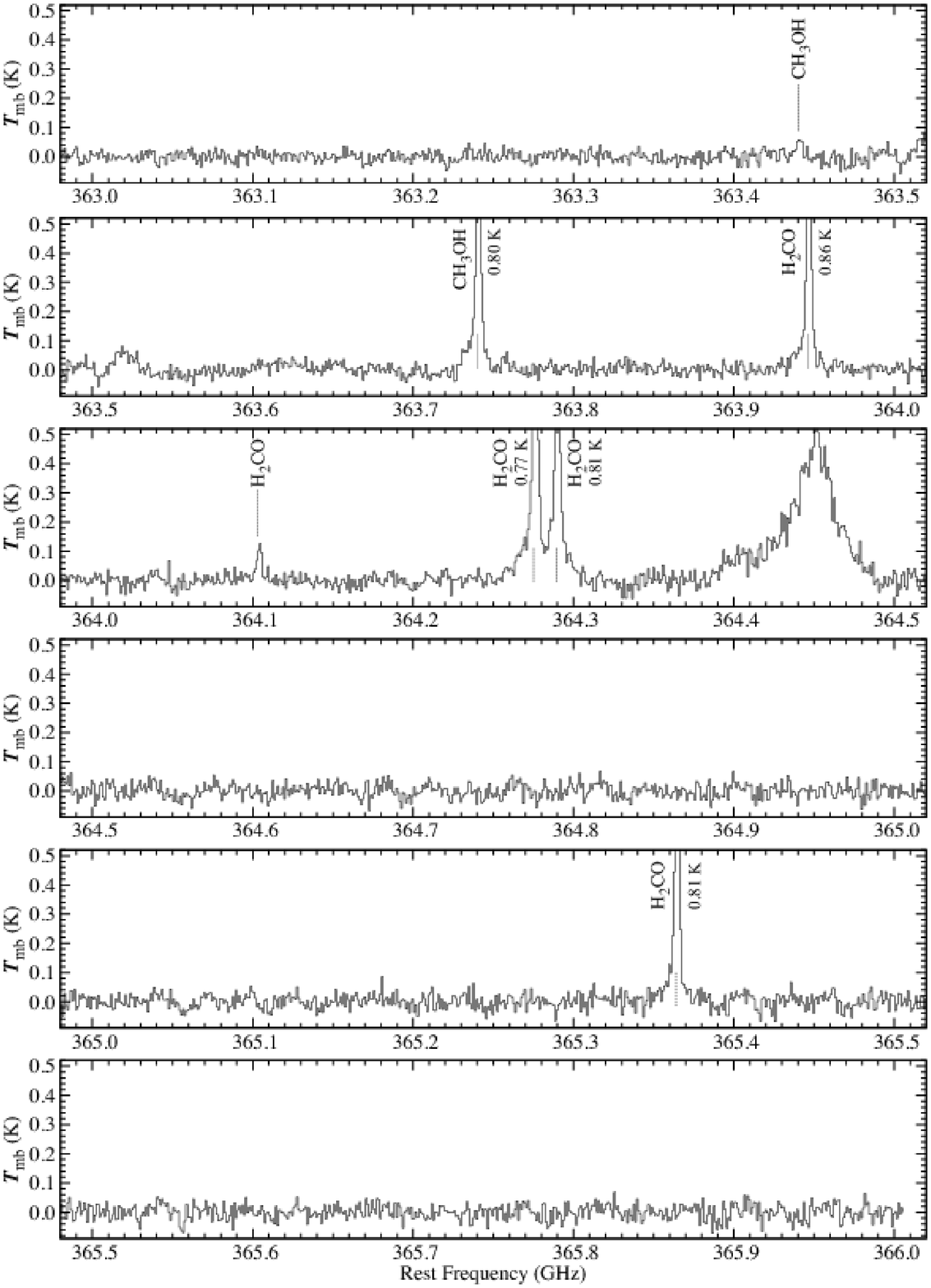}
\caption{\textit{Continued}}
\end{figure}
\setcounter{figure}{1}

\end{document}